\documentclass[amstex]{article}
\usepackage{graphicx}
\usepackage{dcolumn}
\usepackage{amsmath}
\usepackage{amssymb}
\usepackage{amsfonts}
\usepackage{amscd}
\usepackage{bbm} 
\usepackage[usenames]{color}
\newcommand{\QED}{\vrule height5pt width5pt}
\newtheorem{theorem}{Theorem}
\newtheorem{lmm}[theorem]{Lemma}
\newtheorem{cor}[theorem]{Corollary}
\newtheorem{pro}[theorem]{Proposition}
\newtheorem{df}[theorem]{Definition}
\newtheorem{rmk}[theorem]{Remark}

\newcommand\calA{{\cal A}}

\newcommand\calB{{\cal B}}

\newcommand\calK{{\cal K}}

\newcommand\calW{{\cal W}}

\newcommand\bfZ{{\bf Z}}
\newcommand\bfC{{\bf C}}
\newcommand\bfN{{\bf N}}
\newcommand\bfR{{\bf R}}

\newcommand\dprime{{\prime\prime}}
\newcommand\evl{\alpha_t}

\newcommand{\abs}[1]{\left|#1\right|}
\newcommand{\norm}[1]{\left\Vert#1\right\Vert}

\newcommand{\rbk}[1]{\left(#1\right)}
\newcommand{\sbk}[1]{\left[#1\right]}
\newcommand{\cbk}[1]{\left\{#1\right\}}

\newcommand{\innpro}[1]{\left\langle #1 \right\rangle}

\begin{document}
\newpage\thispagestyle{empty}
\begin{center}
{\huge\bf
Regular KMS States 
\\
of 
\\
Weakly Coupled 
\\
Anharmonic Crystals 
\\
and
\\
the Resolvent CCR Algebra}
\\
\bigskip\bigskip
\bigskip\bigskip
{\Large T. Kanda}
\\
{\Large Taku Matsui}
\\
 Graduate School of Mathematics, Kyushu University,
\\
744 Motoka, Nishi-ku, Fukuoka 819-0395, JAPAN
\\
 ma214013@math.kyushu-u.ac.jp
 \\
 matsui@math.kyushu-u.ac.jp
\\
\end{center}
\bigskip\bigskip\bigskip\bigskip
\bigskip\bigskip\bigskip\bigskip
{\bf Abstract:}  
We consider equilibrium states of  weakly coupled anharmonic  quantum oscillators
on $\bfZ$. We consider the Resolvent CCR Algebra introduced by D.Buchholtz and
H.Grundling, and we show that the infinite volume limit of equilibrium states satisfies 
the KMS (Kubo-Martin-Schwinger) condition with regularity(= locally normal to Fock representations). 
 Uniqueness of the KMS states is proven as well. 

\newpage
\section{Introduction}\label{Introduction}
\setcounter{theorem}{0}
\setcounter{equation}{0}
In this article, we consider KMS states of certain one-parameter group of automorphisms 
of ${\rm C}^{*}$-algebra associated with canonical commutation relations (CCR) 
, which ,physically, correspond to  
equilibrium states of  weakly coupled anharmonic  quantum oscillators on $\bfZ$.
\par
Let $(X, \sigma)$ be a 2n dimensional real symplectic vector space with a non-degenerate
symplectic form $\sigma$ and let $\calA_{CCR}( \sigma)$ be the CCR algebra associated with
$(X, \sigma)$ which is a *-algebra of unbounded operators generated by formally self-adjoint
elements $\Psi (f)$ ($f \in X$). $\Psi (f)$ is linear on $f$, 
satisfying CCR
$$
[ \Psi (f), \Psi(g) ]=i \sigma(f, g) \mathbbm{1} , \quad \Psi (f)^* =\Psi (f)
$$
By  the Heisenberg time evolution of a quantum observable $Q$ associated with a Hamiltonian, 
we mean
$$ \evl (Q) = e^{itH} Q e^{-itH}$$
This expression is formal  . 
Both the Hamiltonian and   elements $Q$ of $\calA_{CCR}( \sigma)$
are unbounded operators and we have to specify domains of operators. 
Unless the Hamiltonian is bilinear in Boson operators $\Psi (f)$, it is likely 
that $\evl (Q)$ is not in  $\calA_{CCR}( \sigma)$.
A traditional way to handle $\calA_{CCR}( \sigma)$ is to consider 
unitaries generated by $\Psi (f)$, $W(f) =e^{it \Psi (f)}$.
The ${\rm C}^{*}$-algebra generated by $W(f) $ is called the Weyl CCR algebra.
The Weyl CCR algebra is a simple ${\rm C}^{*}$-algebra and was used in study of
Bose-Einstein condensation of the ideal gas. (c.f.\cite{BratteliRobinsonII})
\par 
 Specialists have realized that the Weyl CCR algebra is not a ${\rm C}^{*}$-algebra suitable  for
scattering theory and to statistical mechanics of interacting Bosons. 
The fundamental problem is (non-)existence of one-parameter group of automorphisms, which
is physically equivalent to existence of the Heisenberg time evolution as automorphisms of a ${\rm C}^{*}$-algebra. 
To be precise, let us consider the case of one degree of freedom. Let $X= \bfR^{2}$ with a symplectic basis
$q,p$ with $\sigma (q,q) = \sigma (p,p)=0$ , $\sigma (q,p) = -\sigma (p,q)=1$. We consider the standard
representation on $L^{2}(\bfR )$, hence $q$ is the multiplication operator and $p$ is the differential operator.
Let $v(q)$ be a potential where $v$ is continuous, with compact support.
Set $h_{0} = \frac{p^{2}}{2}$ and $h=h_{0}+v(q)$.
In \cite{Buchholz2}, D.Buchholz and H.Grundling pointed out that,
for any bounded operator $Q$ on   $L^{2}(\bfR )$,
$$ e^{ith} Q e^{-ith} - e^{ith_{0}} Q e^{-ith_{0}}$$
is a compact operator.
(See also  \cite{Fannes}.)
This implies that any ${\rm C}^{*}$-algebra on $L^{2}(\bfR )$
 invariant  for both the free time evolution and 
 the time evolution with a potential $v$ with a compact support
must contain compact operators. 
This rules out the Wey CCR algebra.
\par
In their research of mathematical foundation of the supersymmetric quantum field theory, 
D.Buchholz and H.Grunrdling introduced a new approach for study of the CCR algebra,
{\em the Resolvent CCR algebra} in  \cite{Buchholz1} (See  \cite{Buchholz2},  \cite{Buchholz3} as well)
The resolvent CCR algebra is a unital ${\rm C}^{*}$-algebra generated by the resolvent of the
field operators
$$R(\lambda , f) =\frac{1}{ i\lambda  +\Psi (f)}$$
where $\lambda$ is a non-zero real parameter. (See \cite{Buchholz2}
for the precise definition of the resolvent CCR algebra.)
\\
The Resolvent CCR algebra has several advantages.  
\\
(i) Singular representations of the CCR algebra with infinite field strength 
can be characterized in terms of the kernel of the semi-resolvent operators. 
\\
(ii) The Fock representation is faithful and there is a one-to-one correspondence between
the regular representations of the Weyl CCR algebra  and those of the resolvent CCR algebra.
\\
(iii) The resolvent CCR algebra for a finite quantum system contains compact operators.
 \par
 Later, we will see the Hamiltonians of weakly coupled anharmonic oscillators 
 gives rise to a one-parameter group of automorphisms on the resolvent CCR algebra and 
 we consider KMS states.
 \par
 To describe precise statements of our results, we introduce notations now. 
 For any positive integer $L$ we set
$$\Lambda_{L} = \left\{ \:\: j \: \in \: \bfZ \:\: \mid  -L < j \leq L \:\:\right\} \subset \bfZ$$
$$\frak H_{\Lambda_{L}} = \otimes_{k \in\Lambda_{L}} L^{2}(\bfR , dx_{k} )$$ 
and let $\calB_{n}$ be the unital abelian ${\rm C}^{*}$-algebra  on $\frak H_{\Lambda_{L}} $ generated by the unit and multiplication operators associated with the functions
 of the form 
 $$f(s_{-L+1}x_{-L+1}+ s_{-L+2}x_{-L+2} \cdots + s_{L}x_{L})$$ 
 where $f(x)$ is a continuous function (with a single real variable) vanishing at infinity
and $s_{-L+1}, \cdots s_{L}$ are real constants.
 Let $\frak R_{L}$ be the unital ${\rm C}^{*}$-algebra  on $\frak H_{\Lambda_{L}}$ generated by the unit and 
 operators 
$$g(s_{-L+1}x_{-L+1} +\cdots + s_{L}x_{L} +t_{-L+1} p_{-L+1}+\cdots + t_{L} p_{L} )$$ 
 where $g(x)$ is a continuous function (with a single variable) vanishing at infinity, $s_{k}, t_{l} ( k, l \in \Lambda_{L} )$ are real constants and
$p_{k}$ be the quantum mechanical momentum operator $p_{k} = -i \frac{\partial}{\partial x_{k}}$.  
Due to the tensor product structure of $\frak H_{\Lambda_{L}}$ we obtain
the natural inclusion $\frak R_{L} \subset \frak R_{M}$  
if $L < M$.
The  inductive limit ${\rm C}^{*}$-algebra of $\cup_{L}\frak R_{\Lambda_{L}}$
is denoted by $\frak R$.
We call $\frak R_{L}$ and $\frak R$ the resolvent CCR algebra.  

 \begin{theorem}
 \label{sec:mainTh1}
 Let $H_{L}$ be the Schr\"{o}dinger operator defined by
\begin{equation}
 \label{eqn:Hamiltonian1}
 H_{L} = \sum_{k= - L+1}^{L}  \left\{\:\:  p_{k}^{2} +  \omega^{2} x_{k}^{2} + V(x_{k})  \:\:\right\}
 +   \sum_{k= - L+1}^{L-1}  \varphi (x_{k} -  x_{k+1}) 
\end{equation}
where the potential $V$  and $\varphi$ are rapidly decreasing smooth functions.
\par
 $\widetilde{\alpha}^{L}_{t}$ defined on the Fock representation via the following equation
 \begin{equation}
\label{eqn:Hamiltonian2}
\widetilde{\alpha}^{L}_{t} (Q) = e^{itH_{L}} Q e^{-itH_{L}} ,\quad  Q \in \frak R_{L}
\end{equation}
 gives rise to a one-parameter group of automorphisms on $\frak R_{L}$.
 \end{theorem}
 Combined with Lieb-Robinson bound techniques on Fock spaces, (c.f. \cite{Koma}, \cite{LiebRobinson}, 
\cite{nachtergaele2005}, \cite{nachtergaele2007}, \cite{NachtergaeleSims2007Dec} ,\cite{nachtergaele2014})
 a ${\rm C}^{*}$-dynamical systems can be introduced for weakly coupled anharmonic oscillators on the infinite lattice $\bfZ$.
 \begin{theorem}
 \label{theorem:TimeEvol1}
 The infinite volume limit
 \begin{equation}
 \label{eqn:Hamiltonian3}
\evl (Q) = \lim_{L \to \infty} \widetilde{\alpha}^{L}_{t} (Q) 
 \end{equation}
 exists in the norm topology of  $\frak R$. 
Let  $H^{{\rm free}}$ be the Hamiltoninan of decoupled oscillators 
\begin{equation}
 \label{eqn:Hamiltonian4}
 H^{{\rm free}} = \sum_{k= - \infty}^{\infty}  
 \left\{\:\:  p_{k}^{2} + \omega^{2}  x_{k}^{2} + V(x_{k})  \:\:\right\} 
\end{equation}
and set $ \evl^{{\rm free}} (Q) = e^{it H^{{\rm free}}} Q e^{- it H^{{\rm free}}}$.
\par
Then, $  \evl \circ \alpha_{-t}^{{\rm free}} (Q)$ and $ \alpha_{-t}^{{\rm free}} \circ \evl (Q)$
are continuous in $t$ for the norm topology of $\frak R$.
 \end{theorem} 
 
 Independently, D.Buchholz obtained  the infinite volume limit of dynamics for more general models
 by different methods in  \cite{Buchholz4}. 
 We believe that application of Lieb-Robinson bound techniques in itself 
 will be useful for more advanced research in future.
 
 The main results of this paper is uniqueness of the regular KMS states of weakly coupled 
quantum oscillators. Uniqueness of the KMS states for one-dimensional quantum systems
is a well-known fact, however, in our case, the time evolution $\evl$ is not norm continuous in $t$.
 $\evl$ is continuous in weak topology on the GNS representations which is locally
 quasi-equivalent to the standard Fock representation. 
We  will see that there exists a infinite volume limit $\psi$ of KMS states 
of finite systems such that the restriction $\psi$ to each finite volume is normal
to the Fock representation and $\psi (Q \evl (R))$ is continuous in $t$.
By regularity of states we mean states locally normal to the standard Fock state. 
(See Definition \ref{df:KMSstate} and Theorem \ref{th:regularKMS}  below .)

\begin{theorem}
 \label{theorem:KMSunique1}
 The regular KMS state associated with the Hamiltonian (\ref{eqn:Hamiltonian1}) exists, and 
  is unique.
 \end{theorem} 
  
Statistical Mechanics of anharmonic crystals with the Hamiltonian (\ref{eqn:Hamiltonian5}) 
defined below has been extensively studied by several people.
(c.f.  \cite{Albeverio}, \cite{Minlos}  and the references therein)
\begin{equation}
 \label{eqn:Hamiltonian5}
H = \sum_{j \in \bfZ^{d}} \{ p^{2}_{j} + V(x_{j} ) \} + \sum_{j, i \in {\bfZ^{d}}, ||i-j|| =1} |x_{i} - x_{j} |^{2}
\end{equation}
where $V$ is a polynomial giving rise to a double well potential.  
Note that, in our anharmonic crystal, Bose particles are fixed on the lattice sites and they are distinguishable.
\par
Results obtained so far are based on perturbation theory
and for developing  a general theory
 a missing point is a suitable ${\rm C}^{*}$-algebra describing full quantum observables.
We believe that the resolvent CCR algebra introduced by D.Buchholz and H.Grunrdling
is the right staff for handling the full quantum  system including momentum operators.
We  hope the results of this article is the first step of  understanding equilibrium states
of  anharmonic crystals. 

\section{Resolvent CCR algebra}\label{sec:Notation}
\setcounter{theorem}{0}
\setcounter{equation}{0}

In this section, we introduce the notation and recall the results of the resolvent CCR algebra in \cite{Buchholz2}.

For a given subset $\Lambda \subset \bfZ$, we denote $c_c(\Lambda)$ by the space of all finitely supported function $f : \Lambda \to \bfC$.
We define the symplectic form $\sigma$ on $c_c(\Lambda)$ by $\sigma(f,g) = {\rm Im}\innpro{f,g}_{\ell^2}$ for $f,g \in c_c(\Lambda)$, where $\innpro{,}_{\ell^2}$ is the canonical inner product on $\ell^2(\bfZ)$.
Then $c_c(\Lambda)$ equipped with $\sigma$ is also a symplectic space.

We consider the Hilbert space ${\frak H}_{\Lambda}$ associated with any finite subset $\Lambda$ of ${\bfZ}$ defined by
\begin{equation*}
{\frak H}_{\Lambda} = \bigotimes_{k \in \Lambda} L^2 ({\bfR}, dx_k ),
\end{equation*}
where $dx_k$ is the Lebesgue measure on ${\bfR}$. 
To simplify the notation, for any finite subsets $\Lambda \subset \Gamma \subset \bfZ$, we identify the linear operator $A$ on ${\frak H}_{\Lambda}$ with the linear operator $A \otimes \mathbbm{1}_{\Gamma \backslash \Lambda}$ on ${\frak H}_{\Gamma}$, where $\mathbbm{1}_{\Gamma \backslash \Lambda}$ is the identity operator on ${\frak H}_{\Gamma \backslash \Lambda}$.
Thus, for any finite subset $\Lambda \subset \bfZ$, we identify the multiplication operator $x_k$ on $L^2(\bfR, dx_k)$, $k \in \Lambda$, and $x_k \otimes \mathbbm{1}_{\Lambda \backslash \{ k \}}$ on ${\frak H}_{\Lambda_L}$.
Also, we identify the differential operator $p_k = - i \frac{\partial}{\partial x_k}$ and $p_k \otimes \mathbbm{1}_{\Lambda \backslash \{ k \}}$.
We denote the trace on ${\frak H}_L$ by ${\rm Tr}_L$.

For any subset $\Lambda$ of $\bfZ$, we denote ${\cal W}(\Lambda)$ and ${\frak R}(\Lambda)$ by the Weyl CCR algebra and the resolvent CCR algebra over $(c_c(\Lambda), \sigma)$, respectively.
The definitions of the Weyl CCR algebra and the resolvent CCR algebra are as follows.

The Weyl CCR algebra is the ${\rm C}^*$-algebra generated by $W(f)$, $f \in c_c(\Lambda)$, satisfying
\begin{eqnarray*}
W(f)^* &=& W(-f), \\
W(f)W(g) &=& e^{-i\frac{ \sigma(f,g) }{2}} W(f+g)
\end{eqnarray*}
for all $f,g \in c_c(\Lambda)$ (see e.g. \cite[Theorem 5.2.8.]{BratteliRobinsonII}.).

The resolvent CCR algebra ${\frak R}(\Lambda)$ is the universal ${\rm C}^*$-algebra generated by $R(\lambda, f)$, $\lambda \in \bfR \backslash \{ 0 \}$, $f \in c_c(\Lambda)$, satisfying
\begin{eqnarray}
R(\lambda, 0) &=& -\frac{i}{\lambda}, \\
R(\lambda, f)^* &=& R(-\lambda, f), \\
\nu R(\nu \lambda, \nu f) &=& R(\lambda, f), \\
R(\lambda, f) - R(\mu, f) &=& i(\mu - \lambda)R(\lambda, f) R(\mu, f), \\
\left[ R(\lambda, f), R(\mu, g) \right] &=& i \sigma (f,g) R(\lambda, f) R(\mu, g)^2 R(\lambda, f), \label{eq:commrelres}\\
R(\lambda, f) R(\mu, g) &=& R(\lambda + \mu, f + g) \{ R(\lambda, f) + R(\mu,g) \nonumber \\
& & + i \sigma(f,g) R(\lambda, f) R(\mu, g) \} \quad \quad (\lambda \not =  -\mu), \label{eq:relres6}
\end{eqnarray}
where $\lambda, \mu, \nu \in \bfR \backslash \{ 0 \}$ and $f,g \in c_c(\Lambda)$.
(See \cite{Buchholz2}.) 
For any subset $\Lambda \subset \bfZ$, the resolvent CCR algebra ${\frak R}(\Lambda)$ is the inductive limit of the net of all finite dimensional non-degenerate symplectic subspaces of $c_c(\Lambda)$ \cite[Theorem 4.9 (ii)]{Buchholz2}.

For any positive integer $L$, we set $\Lambda_L = \{ j \in \bfZ \mid -L < j \leq L \}$. 
For simplicity, we set ${\frak R} = {\frak R}(\bfZ)$ and ${\frak R}_L = {\frak R}(\Lambda_L)$.
Also, we set ${\frak R}_{L^c} = {\frak R}(\bfZ \backslash \Lambda_L)$ and ${\frak R}_{L^\prime \backslash L} = {\frak R}(\Lambda_{L^\prime} \backslash \Lambda_{L})$, for any positive integers $L \leq L^\prime$. 
For the Weyl CCR algebra, we also set ${\cal W} = {\cal W}(\bfZ)$, ${\cal W}_L = {\cal W}(\Lambda_L)$, ${\cal W}_{L^c} = {\cal W}(\bfZ \backslash \Lambda_L)$ and ${\cal W}_{L^\prime \backslash L} = {\cal W}(\Lambda_{L^\prime} \backslash \Lambda_L)$, for any positive integers $L \leq L^\prime$.

Let $\pi_0$ be the Schr\"{o}dinger representation of ${\frak R}(\Lambda)$ on ${\frak H}_{\Lambda}$.
Due to \cite[Theorem 4.10]{Buchholz2}, $\pi_0$ is a faithful representation of ${\frak R}(\Lambda)$.

\section{Lieb-Robinson bounds and limiting dynamics for the resolvent CCR algebra}\label{sec:LiebRobinson}
\setcounter{theorem}{0}
\setcounter{equation}{0}
In this section, we prove the LiebLieb-Robinson bounds for weakly coupled anharmonic quantum oscillators on the resolvent CCR algebra.
First, we introduce the notation.

For any $L \in \bfN$ and positive constant $\omega \geq 0$, let $H^h_L$ be the self-adjoint operator on ${\frak H}_{\Lambda_L}$ defined by
\begin{equation*}
H^h_L = \sum_{k \in \Lambda_L}( p_k^2 +  \omega^2 x_k^2 ).
\end{equation*}
We define the automorphism $\widetilde{\alpha}_t^{h,L}$ on $\calB({\frak H}_{\Lambda_L})$ by
\begin{equation*}
\widetilde{\alpha}_t^{h, L} (Q) = e^{itH_L^h} Q e^{-itH_L^h}, \quad Q \in \calB({\frak H}_{\Lambda_L}).
\end{equation*}
Since the automorphism $\widetilde{\alpha}_t^{h,L}$ induce the symplectic transform on $(c_c(\Lambda_L), \sigma)$, $\widetilde{\alpha}_t^{h,L}$ is an automorphism on $\pi_0({\frak R}_L)$.
Let $\Phi$ be the map from any finite subset $\Lambda$ of $\bfZ$ to ${\cal B}({\frak H}_{\Lambda})$ defined by
\begin{eqnarray}
\Phi(\Lambda) = \left\{
\begin{array}{l}
V(x_k) \quad (\Lambda = \{ k \}) \\
\varphi(x_k - x_{k+1}) \quad (\Lambda = \{ k, k+1 \}) \\
0 \quad (\text{otherwise})
\end{array} \right.
, \label{eq:interaction}
\end{eqnarray}
where $V$ and $\varphi$ are real valued Schwarz functions on ${\bfR}$. 
The function $V$ represent anharmonicity of the potential of the system and $\varphi$ corresponds to the nearest neighbor interaction of particles.
For any finite subset $\Gamma \subset \bfZ$, we set $\Upsilon(\Gamma) = \sum_{\Lambda \subset \Gamma}\Phi(\Lambda)$.
For simplicity, we set $\Upsilon_L = \Upsilon(\Lambda_L)$, $L \in \bfN$ and $\Upsilon_{L^\prime \backslash L} = \Upsilon(\Lambda_{L^\prime} \backslash \Lambda_L)$ whenever $L \leq L^\prime$.
Let $H_L$ be the self-adjoint operator on ${\frak H}_{\Lambda_L}$ defined by
\begin{equation}
H_L = H^h_L + \Upsilon_L = \sum_{k \in \Lambda_L} (p_k^2 + \omega^2 x_k^2 + V(x_k)) + \sum_{k, k+1 \in \Lambda_L} \varphi(x_k - x_{k+1}). \label{eq:anharmonicdynamics}
\end{equation} 
\subsection{Proof of Theorem \ref{sec:mainTh1}}
As the Fock representation is faithful , we consider the convergence in strong and norm topologies  of 
operator valued integral on the Fock space.
Let $U_L(t)$ be the unitary operator on ${\frak H}_{\Lambda_L}$ defined by $U_L(t) = e^{itH_L} e^{-itH^h_L}$.
By using the Dyson series expansion of $U_L(t)$, we obtain
\begin{equation*}
U_L(t) = \mathbbm{1}+\sum_{n \geq 1} i^n \int_0^t dt_1 \int_0^{t_1} dt_2 \cdots \int_0^{t_{n-1}} dt_n \widetilde{\alpha}^{h,L}_{t_n}(\Upsilon_L) \cdots \widetilde{\alpha}^{h,L}_{t_1} (\Upsilon_L).
\end{equation*}
(See e.g. \cite[Theorem 3.1.33]{BratteliRobinsonII}.)
Generally speaking, the above integral makes sense in the weak topology on the Fock space.
However, in our current situation, D.Buchholz and H.Grundling have shown that
$$ \int_{0}^{t}   \widetilde{\alpha}^{h,L}_{t_n} (V(x_{k} ) , \quad 
 \int_{0}^{t}   \widetilde{\alpha}^{h,L}_{t_n} (\varphi (x_{k} -x_{k-1}) $$
are norm continuous families of elements of the resolvent CCR algebra.
(See \cite[Proposition 6.1]{Buchholz2} and \cite[Proof of Proposition 7.1]{Buchholz2})
and as a consequence, the Dyson series converge in the norm topology. 

Thus, $\widetilde{\alpha}_t^L$ is a one-parameter group of automorphism on $\pi_0({\frak R}_L)$.
\QED

\subsection{Lieb-Robinson bounds and limiting dynamics}

Next, we consider the Lieb-Robinson bound of the automorphism $\widetilde{\evl}^L$. 
Before we prove the Lieb-Robinson bound of $\widetilde{\evl}^L$, we introduce the following notations.
(See also \cite{nachtergaele2008}, \cite{nachtergaele2010} and \cite{nachtergaele2014}.)

We set 
\begin{equation*}
H_L^{{\rm free}} = \sum_{k \in \Lambda_L} (p_k^2 + \omega^2 x_k^2 + V(x_k))
\end{equation*}
and
\begin{equation*}
\alpha_t^{{\rm free}, L}(Q) = e^{itH_L^{{\rm free}}} Q e^{-itH_L^{{\rm free}}},  \quad Q \in {\frak R}_L.
\end{equation*}

For any subset $\Gamma \subset \bfZ$ and any finite subset $\Lambda \subset \Gamma$, we set
\begin{equation*}
S_\Gamma(\Lambda) = \{ X \subset \Gamma \mid X \cap \Lambda \not = \emptyset, X \cap (\Gamma \backslash \Lambda) \not = \emptyset \}.
\end{equation*}
If $\Gamma = \bfZ$, then we denote $S(\Lambda)$ by $S_\bfZ(\Lambda)$.
Let $\partial_\Phi \Lambda$ be the subset of $\Lambda$ defined by
\begin{equation*}
\partial_\Phi \Lambda = \{ x \in \Lambda \mid \text{{\rm for some}} X \in S(\Lambda) \text{{\rm with }} x \in X, \Phi(X) \not = 0\}.
\end{equation*}
For any finite subsets $\Gamma_1, \Gamma_2 \subset \bfZ$, we set
\begin{equation*} 
D(\Gamma_1, \Gamma_2) = \min \cbk{ \sum_{x \in \partial_\Phi \Gamma_1} \sum_{y \in \Gamma_2} \frac{1}{1 + \abs{x - y}}, \sum_{x \in \Gamma_1} \sum_{y \in \partial_\Phi \Gamma_2} \frac{1}{1 + \abs{x - y}} }.
\end{equation*}
For $L \in \bfN$, we set $C = 4 \sum_{x \in \bfZ}\frac{1}{(1 + \abs{x})^2}$.
We define the norm $\norm{\cdot}_{{\rm int}}$ of the map $\Phi$ by
\begin{equation*}
\norm{\Phi}_{{\rm int}} = \sup_{\substack{x,y \in \bfZ \\ x \not = y}} \frac{1}{(1 + \abs{x - y})^2} \sum_{\substack{\Lambda: x,y \in \Lambda \subset \bfZ \\ \abs{\Lambda} < \infty}} \norm{\Phi(\Lambda)}
\end{equation*}
where $\abs{\Lambda}$ is the number of elements of $\Lambda$.

\begin{lmm} \label{sec:Lieb-RobinsonBound}
Let $\Gamma_1$ and $\Gamma_2$ be finite disjoint subsets of ${\bfZ}$.
For any finite subset $\Lambda_L$ of $\bfZ$, $L \in \bfN$, with $\Gamma_1 \cup \Gamma_2 \subset \Lambda_L$ and arbitrary $Q \in \pi_0({\frak R}(\Gamma_1))$ and $R \in \pi_0({\frak R}(\Gamma_1))$, it follows that 
\begin{equation}
\norm{ \left[ \widetilde{\alpha}_t^L( \widetilde{\alpha}_{-t}^{ {\rm free}, L } ( Q ) ), R \right] } \leq \frac{2 \norm{Q} \norm{R}}{C} \rbk{e^{2\norm{\Phi}_{ {\rm int} } C \abs{t}} - 1} D(\Gamma_1,\Gamma_2).
\end{equation}
holds for any $t \in {\bfR}$.
\end{lmm}
{\bf Proof.}
Let $\varphi(\Lambda_L) = \sum_{k, k+1 \in \Lambda_L} \varphi(x_{k+1} - x_k)$ and
for any finite subset $\Lambda \subset \bfZ$ let ${\frak Q}(\Lambda)$ be the norm dense subset of $\pi_0({\frak R}(\Lambda))$ defined by
\begin{eqnarray*}
{\frak Q}(\Lambda)= {\rm span}\{ \pi_0(R(\lambda_1, f_1) \cdots R(\lambda_n, f_n)) \mid \lambda_i \in \bfR \backslash \{ 0 \}, \, f_i \in c_c(\Lambda), \, i = 1, \cdots, n, \, n \in \bfN \}.
\end{eqnarray*}
Put
\begin{equation}
F(t) = \sbk{ \widetilde{\alpha}_t^L\rbk{ \widetilde{\alpha}_{-t}^{ {\rm free}, L }( Q ) }, R },
\end{equation}
for $Q \in {\frak Q}(\Gamma_1)$ and $R \in {\frak Q}(\Gamma_2)$.
Since $R(\lambda, f)$ preserves the domain of multiplication and differential operators by \cite[Theorem4.2 (i)]{Buchholz2} and the Schwarz function in ${\frak H}_{\Lambda_L}$ is analytic elements for the operator $\sum_{k \in \Lambda_L} (p_k^2 + \omega^2 x_k^2 + V(x_k))$, the function $F$ is strongly differentiable.
Thus, the derivation of $F$ is
\begin{equation}
\frac{d}{dt} F (t) = i \sbk{ \widetilde{\alpha}_t^L ( \varphi(\Lambda_L) ), F(t) } - i \sbk{ \widetilde{\alpha}_t^L \rbk{ \widetilde{\alpha}_{-t}^{ {\rm free}, L }(Q) },\sbk{ \widetilde{\alpha}_t^L(\varphi(\Lambda_L)), R } }. \label{eq:diffofF}
\end{equation}
Since $\widetilde{\alpha}_t^L$ is strongly continuous on ${\frak H}_{\Lambda_L}$, the solution $F(t)$ of the equation (\ref{eq:diffofF}) satisfies the following estimate by \cite[Lemma 2.2]{nachtergaele2014}:
\begin{eqnarray*}
\norm{F(t)} \leq \norm{F(0)} + \int_0^{\abs{t}} ds \, \norm{ \sbk{ \widetilde{\alpha}_s^L \rbk{ \widetilde{\alpha}_{-s}^{ {\rm free}, L }(Q) },\sbk{ \widetilde{\alpha}_s^L(\varphi(\Lambda_L)), R } }}.
\end{eqnarray*}
Since for any subset $\Lambda \subset \bfZ$, ${\frak Q}(\Lambda)$ is a norm dense subset in $\pi_0({\frak R}(\Lambda))$, the above inequality holds for any elements of $Q \in \pi_0({\frak R}(\Gamma_1))$ and $R \in \pi_0({\frak R}(\Gamma_2))$.

By the proof of \cite[Theorem 3.1]{nachtergaele2014}, we get
\begin{equation}
\norm{ \left[ \widetilde{\alpha}_t^L( \widetilde{\alpha}_{-t}^{ {\rm free}, L } ( Q ) ), R \right] } \leq \frac{2 \norm{Q} \norm{R}}{C} \rbk{e^{2\norm{\Phi}_{ {\rm int} } C \abs{t}} - 1} D(X,Y).
\end{equation}
for any $Q \in \pi_0({\frak R}(\Gamma_1))$ and $R \in \pi_0({\frak R}(\Gamma_2))$.
\QED

We define the automorphisms $\alpha_t^L$ and $\alpha_t^{h, L}$ ($t \in \bfR$) on ${\frak R}_L$ by
\begin{eqnarray} \label{eq:anharmonicbddauto}
& & \alpha_t^L(Q) = \pi_0^{-1}(e^{itH_L}) Q \pi_0^{-1}(e^{-itH_L}), \\
& & \alpha_t^{{\rm free},L}(Q) = \pi_0^{-1}(e^{itH^{{\rm free}}_L}) Q \pi_0^{-1}(e^{-itH^{{\rm free}}_L})
\end{eqnarray}
for $Q \in {\frak R}_L$.
Note that the Schr\"{o}dinger representation $\pi_0$ is faithful representation.
Thus, we get the following.

\begin{cor}
Let $\Gamma_1$ and $\Gamma_2$ be finite disjoint subsets of ${\bfZ}$.
For any finite $\Lambda_L$ with $\Gamma_1 \cup \Gamma_2 \subset \Lambda_L$ and arbitrary $Q \in {\frak R}(\Gamma_1)$ and $R \in {\frak R}(\Gamma_1)$, it follows that 
\begin{equation}
\norm{ \left[ \evl^L( \alpha_{-t}^{ {\rm free}, L } ( Q ) ), R \right] } \leq \frac{2 \norm{Q} \norm{R}}{C} \rbk{e^{2\norm{\Phi}_{ {\rm int} } C \abs{t}} - 1} D(\Gamma_1,\Gamma_2).
\end{equation}
holds for all $t \in {\bfR}$.
\end{cor}

\begin{theorem} \label{sec:existenceautomorphism}
For any $t \in \bfR$, $L \in \bfN$ and $Q \in {\frak R}_L$, the norm limit
\begin{equation}
\lim_{N \rightarrow \infty} \evl^{N} ( Q ) = \evl(Q)
\end{equation}
exists and the convergence is uniform for $t$ in compact sets.
\end{theorem}

\noindent
{\bf Proof.} The assertion follows from the proof of \cite[Theorem4.1]{nachtergaele2014} and the above corollary. \QED

Finally, we give the proof of Theorem \ref{theorem:TimeEvol1}.

\subsection{Proof of Theorem \ref{theorem:TimeEvol1}.}
The existence of the infinite volume limit is proven in Theorem \ref{sec:existenceautomorphism}.
Thus, we prove the continuity of $\alpha_t \circ \alpha_{-t}^{{\rm free}}$ in $t \in \bfR$.
We may assume that $t \in \sbk{ 0, T }$.
By the Dyson series expansions of $e^{itH_L} e^{-itH_L^{{\rm free}}}$, we obtain
\begin{eqnarray}
\norm{e^{itH_L} e^{-itH_L^{{\rm free}}} - \mathbbm{1}} &=& \norm{\sum_{n \geq 1} i^n \int_0^{t} dt_1 \cdots \int_0^{t_{n-1}} dt_n \alpha_{t_n} (\varphi(\Lambda_L)) \cdots\alpha_{t_1}(\varphi(\Lambda_L))} \nonumber \\
&\leq& e^{2 L \norm{\varphi}_\infty T } - 1, \label{eq:estimateDyson}
\end{eqnarray}
where $\varphi(\Lambda_L) = \sum_{k, k+1 \in \Lambda_L} \varphi(x_{k+1} - x_k)$.
Note that $\norm{\Phi}_{{\rm int}} \leq \frac{1}{4} \norm{\varphi}_\infty$, where $\norm{\varphi}_\infty$ is the supremum norm of $\varphi$.
By using the estimate (83) of the proof of \cite[Theorem 4.1]{nachtergaele2014}, for any $L \in \bfN$, $L \leq N \leq N^\prime$ and $Q \in {\frak R}_L$, we have
\begin{eqnarray*}
& & \norm{\alpha_t^{N^\prime} \circ \alpha_t^{{\rm free}, N^\prime} (Q) - \alpha_t^{N} \circ \alpha_t^{{\rm free}, N} (Q) } \\
&\leq& \frac{1}{2}T(1 + e^{\frac{1}{2} C \norm{\varphi}_\infty T}) \norm{Q} \sum_{k \in \Lambda_L} \sum_{l \in \Lambda_{N^\prime} \backslash \Lambda_N} \frac{1}{(1 + \abs{k-l})^2} .
\end{eqnarray*}
Thus, when $N = L$ and $N^\prime \to \infty$, we obtain
\begin{equation}
\norm{\alpha_t \circ \alpha_t^{{\rm free}} (Q) - \alpha_t^{L} \circ \alpha_t^{{\rm free}, L} (Q) } \leq T(1 + e^{\frac{1}{2} C\norm{\varphi}_\infty T})\norm{Q}LC. \label{eq:estimateauto}
\end{equation}
By (\ref{eq:estimateDyson}) and (\ref{eq:estimateauto}), it follows that
\begin{eqnarray*}
\norm{\alpha_t \circ \alpha_t^{{\rm free}}(Q) - Q} &\leq& \norm{\alpha_t \circ \alpha_t^{{\rm free}} (Q) - \alpha_t^{L} \circ \alpha_t^{{\rm free}, L} (Q) } \\
& & + 2\norm{e^{itH_L} e^{-itH_L^{{\rm free}}} - \mathbbm{1}} \norm{Q} \\
&\leq& T(1 + e^{\frac{1}{2} C\norm{\varphi}_\infty T})\norm{Q}LC + 2(e^{2TL\norm{\varphi}_\infty} - 1)\norm{Q}. 
\end{eqnarray*}
Thus, we are done.\QED


\section{Regular states of the resolvent CCR algebra}\label{sec:regular}
\setcounter{theorem}{0}
\setcounter{equation}{0}
In this section, we consider regular states on ${\frak R}_L$, $L \in \bfN$, or ${\frak R}$.
Recall that a state $\psi$ on ${\frak R}_L$ or ${\frak R}$ is regular, if and only if $\ker (\pi_\psi (R(\lambda, f))) = \{ 0 \}$
 for any $\lambda \in {\bfR} \backslash \{ 0 \}$ and $f \in c_c(\Lambda_L)$ or $f \in c_c(\bfZ)$, respectively, where $\pi_\psi$ is the GNS representation associated with  $\psi$. ( c.f.   \cite[Definition 4.3]{Buchholz2} )
 In another word , $\pi_\psi (R(\lambda, f))$ is the resolvent of a closed operator if $\psi$ is regular.
Note that there is a one-to-one correspondence between a regular state of ${\frak R}$ and that of 
${\cal W}$. (See \cite[Corollary 4.4.]{Buchholz2} .) and by abuse of notations,  we employ the same notation , $\psi$ or $\varphi$
etc. for the regular states of ${\frak R}$ and ${\cal W}$.

The following claims are straight forward implication of the Stone-von Neumann uniqueness theorem.
 (See e.g. \cite[Corollary 5.2.15]{BratteliRobinsonII}.)
\begin{lmm}
Let $\psi$ be a regular state of ${\frak R}_L$.
Then. $\psi$ is normal with respect to the Fock representation.
\end{lmm}



\begin{cor} \label{sec:regstate}
Let $\psi$ be a regular state on ${\frak R}_L$.
Then there exists a positive trace class operator $\rho$ on ${\frak H}_{\Lambda_L}$ such that ${\rm Tr}_L (\rho) = 1$ and $\psi(Q) = {\rm Tr}_L(\rho \pi_0(Q))$, where $\pi_0$ is the Schr\"{o}dinger representation of ${\frak R}_L$ and ${\rm Tr}_L$ is the trace on ${\frak H}_{\Lambda_L}$.
\end{cor}

\begin{lmm} \label{sec:sdensecpt}
Let $\psi$ be a regular state on ${\frak R}_L$, $L \in \bfN$.
Let $({\frak H}_\psi, \pi_\psi, \xi_\psi)$ be the GNS representation of $\psi$.
Put ${\frak K}(\Lambda_L) = \pi_0^{-1}({\cal K}({\frak H}_{\Lambda_L}))$, where ${\cal K}({\frak H})$ is the set of all compact operator on a Hilbert space ${\frak H}$.
Then, $\pi_\psi({\frak K}(\Lambda_L))$ is weekly dense in $\pi_\psi({\frak R}_L)^{\dprime}$.
\end{lmm}
\begin{pro} \label{sec:continuityofregularstate}
Let $\psi$ be a regular state on ${\frak R}_L$, $L \in \bfN$.
Then, $\psi(\alpha_t^L(Q)R)$ is continuous on $t \in \bfR$ for any $Q, R \in {\frak R}_L$ where $\alpha^L_t$ is defined in {\rm (\ref{eq:anharmonicbddauto})}.
\end{pro}

\noindent
{\bf Proof.} 
In ${\cal B}({\frak H}_{\Lambda_L})$, we consider the Dyson series of $U_L(t) = e^{itH_L} e^{-itH^h_L}$ and $U_L(t) - \mathbbm{1}$ has the following estimate:
\begin{eqnarray*}
\norm{U_L(t) - \mathbbm{1}} &=& \norm{ \sum_{n \geq 1} i^n \int_0^{t} dt_1 \cdots \int_0^{t_{n-1}} dt_n \widetilde{\alpha}^{h, L}_{t_n} (\Upsilon_L) \cdots \widetilde{\alpha}^{h, L}_{t_1}(\Upsilon_L)}\\
&\leq& (e^{\abs{t}\norm{\Upsilon_L}} - 1).
\end{eqnarray*}
Note that $\widetilde{\alpha}_t^{h,L}$ preserve the set of all of compact operators ${\cal K}({\frak H}_{\Lambda_L})$ and for $Q \in {\cal K}({\frak H}_{\Lambda_L})$, $\widetilde{\alpha}_t^{h,L}(Q)$ is norm continuous.
Since for $Q \in \mathcal{K}({\frak H}_{\Lambda_L})$, we obtain
\begin{eqnarray}
\norm{e^{itH_L} Q e^{-itH_L} - Q} &=& \norm{U_L(t) e^{itH^h_L} Q e^{-itH^h_L} U_L(t)^{-1} - Q} \nonumber \\
&\leq&  2(e^{\abs{t} \norm{\Upsilon_L}} - 1) \norm{Q} + \norm{\widetilde{\alpha}_t^{h,L}(Q) - Q}. \label{eq:Dysonseriesestimate}
\end{eqnarray}
Since the Schr\"{o}dinger representation $\pi_0$ is faithful, for any $Q \in {\cal K}({\frak H}_{\Lambda_L})$, $\widetilde{\alpha}_t^L(Q)$ is norm continuous for $t \in \bfR$.
By Corollary \ref{sec:regstate} and Lemma \ref{sec:sdensecpt}, $\psi(R\alpha_t^L(Q))$ is continuous for $R, Q \in {\frak R}_L$.
\QED

Next let us  recall the definition of quasi-containment.
Let $\calA$ be a C$^*$-algebra and let $({\frak H}_1, \pi_1)$ and $({\frak H}_2, \pi_2)$ be nondegenerate representations of $\calA$.
The representations $\pi_1$ and $\pi_2$ is quasi-equivalent, if there exists an isomorphism $\gamma : \pi_1(\calA)^\dprime \mapsto \pi_2(\calA)^\dprime$ such that $\gamma(\pi_1(A)) = \pi_2(A)$ for all $A \in \calA$ (see also \cite[Definition 2.4.25]{BratteliRobinsonII} and \cite[Theorem2.4.26]{BratteliRobinsonII}).
If a subrepresentation of $\pi_1$ is quasi-equivalent to $\pi_2$, then $\pi_1$ is quasi-contain $\pi_2$.
The next lemma is essentially due to \cite[Lemma 1]{Araki1975}.

\begin{lmm} \label{sec:quasicontain}
Let $\psi_1$ and $\psi_2$ be regular states on ${\frak R}$ and $({\frak H}_1, \pi_1, \xi_1)$ and $({\frak H}_2, \pi_2, \xi_2)$ be the GNS representations associated with $\psi_1$ and $\psi_2$, respectively.
If $\pi_1$ does not quasi-contain $\pi_2$, then there exists a sequence of projections $e_m \in \bigcup_{L \in \bfN} {\frak R}_L$ such that
\begin{eqnarray}
\lim_m \psi_1(e_m) = 0, \\
\lim_m \psi_2 (e_m) = a > 0.
\end{eqnarray}
\end{lmm}

\noindent
{\bf Proof.}
Put ${\frak H} = {\frak H}_1 \oplus {\frak H}_2$, $\pi = \pi_1 \oplus \pi_2$, $\widetilde{\xi}_1 = \xi_1 \oplus 0$, $\widetilde{\xi}_2 = 0 \oplus \xi_2$ and ${\frak M} = \pi({\frak R})^{\dprime}$.
Note that $\pi$ is a regular representation of ${\frak R}$.
Let $E_1$ and $E_2$ be the projections from ${\frak H}$ onto ${\frak H}_1$ and ${\frak H}_2$, respectively.
By the assumption, there exists a central projection $E \in {\frak M} \cap {\frak M}^\prime$ such that $EE_1 = 0$, $E \leq E_2$.
By Kaplansky's density theorem, there exists self-adjoint elements $a_n \in {\frak R}_{L(n)}$ such that
\begin{equation*}
\mathop{{\rm st}\text{-}\lim}_{n} \pi(a_n) = E,
\end{equation*}
where the $\mathop{{\rm st}\text{-}\lim}$ is the strong limit in ${\frak H}$.
By the regularity of $\pi$, Lemma \ref{sec:sdensecpt} and Kaplansky's density theorem, there exists self-adjoint elements $b_m^{(n)} \in {\frak K}(\Lambda_{L(n)})$ such that
\begin{equation*}
\mathop{{\rm st}\text{-}\lim}_{m} \pi(b^{(n)}_m) = \pi(a_n). 
\end{equation*}
Note that the spectral projections of $b^{(n)}_m$ are also contained in ${\frak R}_L$.
Let $e_n$ be the spectral projection of $b_m^{(n)}$ for an interval $[1 - \delta, 1+\delta]$, where $\delta \in (0,1)$ is fixed.
Then, $e_n \in {\frak K}(\Lambda_{L(n)}) \subset {\frak R}_{L(n)}$ and 
$\mathop{{\rm st}\text{-}\lim}_n \pi(e_n) = E.$
\par
Thus, $\lim_n \psi_1(e_n) = 0, \quad \lim_n \psi_2(e_n) = a > 0. $
\hspace{\fill}\QED

\subsection{Relative entropy}
In this subsection, we recall the definition and the properties of the relative entropy of normal states of von Neumann algebras.

The relative entropy for positive normal linear functionals on a von Neumann algebra was introduced by H. Araki in \cite{ArakiRelEntI} and \cite{ArakiRelEntII}.
Let $\psi_1$ and $\psi_2$ be positive normal linear functionals over a von Neumann algebra ${\frak M}$.
Due to the theory of standard form of von Neumann algebra, there exists a Hilbert space ${\frak H}$ and $\xi_1, \xi_2 \in {\frak H}$ such that $\psi_1(a) = \innpro{\xi_1, a \xi_1}$ and $\psi_2(a) = \innpro{\xi_2, a \xi_2}$, $a \in {\frak M}$.

Let  $S_{\xi_2, \xi_1}$ be the closable densely defined conjugate linear operator $S_{\xi_2, \xi_1}$ defined by
\begin{equation*}
S_{\xi_2, \xi_1} a \xi_1 = a^* \xi_2,
\end{equation*}
for $a \in {\frak M}$.
The relative modular operator $\Delta_{\xi_2, \xi_1}$ is , by definition,
\begin{equation}
\Delta_{\xi_2, \xi_1} = S_{\xi_2, \xi_1}^* \overline{S_{\xi_2, \xi_1}},
\end{equation}
where $\overline{S_{\xi_2, \xi_1}}$ is the closure of the operator $S_{\xi_2, \xi_1}$.
We denotes the projection onto $\overline{{\frak M}^{\prime} \xi_1}$ snd $\overline{{\frak M}^\prime \xi_2}$ on ${\frak H}$ by $s(\psi_1)$ and $s(\psi_2)$, respectively.
We define the relative entropy $S_A(\psi_1, \psi_2)$ of $\psi_1$ and $\psi_2$ by
\begin{equation}
S_A(\psi_1, \psi_2) = \left\{
\begin{array}{ll}
- \int_0^\infty \log (\lambda) d(\xi_1, E(\lambda) \xi_1), & \text{ if } (s(\psi_1) \leq s(\psi_2)) \\
\infty & \text{otherwise}
\end{array}
\right. .
\end{equation}

A. Uhlmann introduced the relative entropy for positive linear functionals on a (not necessarily normed)
 $*$-algebra. (c.f. \cite{Uhlmann}.)
We denote the relative entropy defined by A.Uhlmann by $S_U$.
The definition of relative entropy of A, Uhlmann and that of H.Araki  coincide
 for any positive normal linear functionals on any von Neumann algebra (see \cite{Hiai}.).
We recall that basic properties of relative entropy.
\begin{lmm} {\rm (\cite[Lemma 3.1]{Hiai})} \label{sec:Calgrelent}
Let ${\cal A}$ be a unital ${\rm C}^*$-algebra and $\pi$ be a non-degenerate representation of ${\cal A}$ on a Hilbert space.
If $\psi_1$ and $\psi_2$ are positive linear functionals of ${\cal A}$ with normal extensions $\widehat{\psi_1}$ and $\widehat{\psi_2}$ to $\pi({\cal A})^{\dprime}$ such that $\psi_1(A) = \widehat{\psi_1}(\pi(A))$ and $\psi_2(A) = \widehat{\psi_2}(\pi(A))$, $A \in {\cal A}$,
then 
\begin{equation}
S_U(\psi_1, \psi_2) = S_A(\widehat{\psi_1}, \widehat{\psi_2}).
\end{equation}
\end{lmm}

Thus, in this paper, by the relative entropy of states on a unital C$^*$-algebra, we mean the relative
entropy of normal extension of states to the von Neumann algebra associated with the GNS representation . More precisely,
let $\calA$ be a ${\rm C}^*$-algebra and $\pi$ be a non-degenerate representation of $\calA$ on a Hilbert space ${\frak H}$.
Let $\widehat{\psi_1}$ and $\widehat{\psi_2}$ be positive normal linear functionals on $\pi(\calA)^\dprime$.
We set $\psi_1(A) = \widehat{\psi_1}(\pi(Q))$ and $\psi_2(A) = \widehat{\psi_2}(\pi(Q))$ for $Q \in \calA$.
The relative entropy $S(\psi_1, \psi_2)$ of $\psi_1$ and $\psi_2$ is defined by
\begin{equation*}
S(\psi_1, \psi_2) = S_A(\widehat{\psi_1}, \widehat{\psi_2}).
\end{equation*}

\begin{lmm} \label{sec:restrictionstate}
Let $\psi_1$ and $\psi_2$ be a regular states on ${\frak R}$.
Then 
\begin{equation}
S(\psi_1 \restriction_{{\frak R}_L}, \psi_2 \restriction_{{\frak R}_L}) = S(\widehat{\psi_1} \restriction_{ \pi_0({\frak R}_L)^{\dprime} }, \widehat{\psi_2} \restriction_{ \pi_0({\frak R}_L)^{\dprime} } ).
\end{equation}
\end{lmm}
\noindent
{\bf Proof.}
By Corollary \ref{sec:regstate}, there exists trace class operators $\rho_1$ and $\rho_2$ on ${\frak H}_{\Lambda_L}$.
Thus, we set $\widehat{\psi_1}(\pi_0(Q)) = {\rm Tr}_L(\rho_1 \pi_0(Q))$ and $\widehat{\psi_2}(\pi_0(Q)) = {\rm Tr}_L(\rho_2 \pi_0(Q))$ for $Q \in {\frak R}_L$.
Then, we obtain
\begin{eqnarray*}
S(\psi_1 \restriction_{ {\frak R}_L } , \psi_2 \restriction_{{\frak R}_L} ) &=& S(\widehat{\psi_1} \circ \pi_0 \restriction_{{\frak R}_L}, \widehat{\psi_2} \circ \pi_0 \restriction_{{\frak R}_L}) \\
&=& S(\widehat{\psi_1} \restriction_{\pi_0({\frak R}_L)^{\dprime}}, \widehat{\psi_2} \restriction_{\pi_0({\frak R}_L)^{\dprime}} ).
\end{eqnarray*}
\hspace{\fill}\QED

\begin{lmm} \label{sec:monotonerelentropy} {\rm (\cite[Corollary 5.12 (iii)]{Ohya})}
Let ${\frak N} \subset {\frak M}$ be von Neumann algebras and $\psi_1$ and $\psi_2$ be normal states on ${\frak M}$.
Assume there exists a norm one projection from ${\frak M}$ to ${\frak N}$.
Then 
\begin{equation}
0 \leq S( \psi_1 \restriction_{{\frak N}}, \psi_2 \restriction_{{\frak N}}  ) \leq S(\psi_1, \psi_2).
\end{equation}
\end{lmm}
By the  same argument as that in  \cite{Araki1975}, we have the following.
\begin{lmm}
Let $\psi_1$ and $\psi_2$ be regular states on ${\frak R}$. If 
\begin{equation}
\sup_{L \in \bfN} S(\psi_1 \restriction_{{\frak R}_L } , \psi_2 \restriction_{{\frak R}_L} ) \equiv \mu < \infty,
\end{equation}
then $\pi_2$ quasi contains $\pi_1$ where $\pi_j$ is the GNS representation of ${\frak R}$ associated with $\psi_j$, $j = 1,2$.
\end{lmm}

\noindent
{\bf Proof.} 
Assume that $\pi_2$ does not quasi-contain $\pi_1$. By Lemma \ref{sec:quasicontain}, there exists a sequence of projections $e_n \in {\frak R}_{L(n)}$ such that
\begin{eqnarray*}
\lim_n \psi_1(e_n) = a > 0,\\
\lim_n \psi_2(e_n) = 0.
\end{eqnarray*}
Then,
\begin{equation*}
- \psi_1(e_n) \log \psi_2(e_n) \rightarrow \infty.
\end{equation*}
Consider the ${\rm C}^*$-subalgebra ${\cal B}_n$ of ${\frak R}_{L(n)}$ generated by $e_n$ and $1 - e_n$.
By Lemma \ref{sec:restrictionstate} and Lemma \ref{sec:monotonerelentropy},
\begin{eqnarray*}
S(\psi_1 \restriction_{{\frak R}_L}, \psi_2 \restriction_{{\frak R}_L}) &=& S(\widehat{\psi_1} \restriction_{\pi_0({\frak R}_{L(n)})^\dprime}, \widehat{\psi_2} \restriction_{\pi_0({\frak R}_{L(n)})^\dprime}) \\
&\geq& S(\widehat{\psi_1} \restriction_{\pi_0({\cal B}_n)^\dprime}, \widehat{\psi_2} \restriction_{\pi_0({\cal B}_n)^\dprime}) \\
&=& \psi_1(e_n) \log \frac{\psi_1(e_n)}{\psi_2(e_n)} + \psi_1(1 - e_n) \log \frac{\psi_1(1 - e_n)}{\psi_2(1 - e_n)}.
\end{eqnarray*}
The above estimate contradict to the assumption.
\QED

Finally, we recall the continuity of the relative entropy.

\begin{lmm} {\rm (\cite[Corollary 5.12 (i)]{Ohya})} \label{sec:wcontiofrelent}
Let $\psi_i, \psi, \phi_i$ and $\phi$ be normal states on a von Neumann algebra ${\frak M}$.
If $\psi_i$ and $\phi_i$ converge to $\psi$ and $\phi$ in $\sigma({\frak M}_*, {\frak M})$ topology, respectively, then 
\begin{equation}
S(\psi, \phi) \leq \liminf_{i} S(\psi_i, \phi_i).
\end{equation}
\end{lmm}

\section{KMS states on the resolvent CCR algebra} \label{sec:KMS}
\setcounter{theorem}{0}
\setcounter{equation}{0}
In this section, we consider KMS states on the resolvent CCR algebra.


In our model, the time evolution $\evl (Q)$ is not be norm continuous 
as a function of $t$ for certain $Q$.
However, the set of elements $Q$ for which $\evl (Q)$ have analytic extension as functions of $t$ 
is weakly dense in regular representations. 
We introduce the notion of KMS states in the following manner.

\begin{df}
\label{df:KMSstate}
Let $\evl$, $t \in {\bfR}$, be a (not necessarily continuous) one-parameter group of 
$*$-automorphism on a unital {\rm C}$^*$-algebra ${\cal A}$.
The state $\psi$ is an $(\alpha, \beta)$-KMS state, if $\psi$ is 
$\alpha$ invariant state, {\rm i.e.} $\psi(\evl(Q)) = \psi(Q)$, $Q \in {\cal A}$, and
$\psi(Q \evl(R))$ is a continuous function in $t \in {\bfR}$ for any $Q, R \in {\cal A}$ satisfying the KMS boundary condition, namely, 
there exists a function $F_{Q,R}(t)$
holomorphic in  $I_\beta$, bounded continuous on the closure of $I_\beta$ 
 such that
\begin{equation}
F_{Q,R}(t) =  \psi(Q \evl(R)) , \quad F_{Q,R}(t+i \beta ) =  \psi(\evl(R)Q)
\end{equation}
for any $Q, R \in {\cal A}$.
\end{df}

\subsection{KMS state associated with the weakly coupled anharmonic oscillators}
In this subsection, we consider KMS states on the resolvent CCR algebra associated with the anharmonic dynamics defined in (\ref{eq:anharmonicdynamics}).

To begin with, we recall for our resolvent CCR algebra there exists the trivial (not interesting) state   
$\psi_{trivial}$ defined by
\begin{equation}
 \psi_{trivial} ( R(\lambda_{1} , f_{1})R(\lambda_{2} , f_{2})\cdots R(\lambda_{k} , f_{k}))
=0
\label{eqn:singularKMS10001}
\end{equation}
for any $\lambda_{i} \in \bfC$ and $f_{j}$.
\par
For any finite system, we have decomposition of the KMS state into the regular part
 and the singular part.
 \begin{lmm} \label{lmm:RegularIregularKMSstate}
 We identify ${\frak R}_L$ with operators in the Schr$\ddot{o}$dinger representation 
 on $\frak H_{\Lambda_{L}}= L^{2}(\bfR^{2L})$ and  Let $\calK_{L}$ be the algebra of compact operators on
 $L^{2}(\bfR^{2L})$ which we regard as a sub-algebra of ${\frak R}_L$ . 
 Let $H$ be a positive self-adjoint operator on $\frak H_{\Lambda_{L}}$ 
 satisfying the following conditions. :
 \\ 
 (a) $e^{it H} \pi_{\Lambda_{L}} (Q) e^{-it H}$  ($Q \in {\frak R}_L$) gives rise to a one-parameter 
 group of automorphisms of $ {\frak R}_L$ denoted by $\evl (Q)$.  
 $\pi_{\Lambda_{L}} (\evl (Q))=  e^{it H} \pi_{\Lambda_{L}} (Q) e^{-it H}$ 
 \\
 (b) $e^{-\beta H}$ is a trace class operator on $\frak H_{\Lambda_{L}}$.
 \\
 Let $\psi_{\beta}$ be a $\beta$-KMS state for  $\evl$. There exist 
 $\beta$-KMS states $\psi_{s}$ and $\psi_{r}$
satisfying the following properties. 
\\
(i) The kernel of the GNS representation for $\psi_{s}$ contains the compact operator algebra
$\calK_{L}$ on $\frak H_{\Lambda_{L}}$. 
\\
(ii) $ \psi_{r}$ is the regular KMS state defined by
\begin{equation}
 \psi_{r}(Q) =\frac{tr_{\frak H_{\Lambda_{L}}}(e^{-\beta H} Q)}{ tr_{ \frak H_{\Lambda_{L} } } (e^{-\beta H})} ,
\quad Q \in  {\frak R}_L .
\label{eqn:regularKMS10000}
\end{equation}
(iii) $\psi_{\beta}$ is a convex combination of $\psi_{s}\psi_{s}$ and $\psi_{r}$,
$$ \psi_{\beta} = \lambda \psi_{r} + (1-\lambda )\psi_{s}$$
for some a positive real number $\lambda$ $0\leq \lambda \leq 1$ 
\end{lmm}
\noindent
{\bf Proof.} 
Let $p_{j}$ ($j=0,1,2,\cdots$) be the mutually orthogonal projections in ${\frak R}_L$
such that $\pi_{\Lambda_{L}} (p_{j} )$ is the rank one projection associated with
an eigenvector for an eigenvalue $\epsilon_{j}$ of $H$ and 
$$H = \sum_{j} \epsilon_{j} \pi_{\Lambda_{L}} (p_{j} ) .$$
Set $P_{m} = \sum_{j=1}^{n} p_{j}$ and $P = w-lim_{m \to \infty}\pi_{\psi_{\beta}}(P_{m})$
on the GNS representation associated with $\psi_{\beta}$.
We claim that the projection $P$ is in the centre of the von Neumann algebra $\pi_{\psi_{\beta}}({\frak R}_L )^{\dprime}$.
In fact, by definition $P$ commutes with any elements in $\pi_{\psi_{\beta}}(\calK_{L})$ and for any $Q \in {\frak R}_L$
$QP_{m}$ is of finite rank in the Schr$\ddot{o}$dinger representation, $\pi_{\psi_{\beta}}(QP_{m})$
and its weak limit commutes with $\pi_{\psi_{\beta}}({\frak R}_L )$.
\par
Set $\lambda = \lim_{m\to\infty}\psi (P_{m})$ and 
$$\psi_{r}(Q) = \lim_{m\to\infty} \psi_{\beta}(QP_{m}) , \quad
\psi_{s}(Q) = \lim_{m\to\infty} \psi_{\beta}(Q(1-P_{m})) $$
for any $Q \in {\frak R}_L$.
As $P$ is in the centre of $\pi_{\psi_{\beta}}({\frak R}_L )^{\dprime}$, 
$\psi_{r}$ and $\psi_{s}$ are $\beta -$KMS states.
\par
For any compact $Q \in \calK_{L}$, $\psi_{s}(Q^{*}Q) = \lim_{m\to\infty} \psi_{\beta}(Q^{*}Q(1-P_{m}))=0$ 
as $\{ P_{m}\}$ is an approximate unit for $Q \in \calK_{L}$. As the GNS vector of the KMS state $\psi_{s}$
is separating for $\pi_{\psi_{s}}( {\frak R}_L)$ the kernel of $\pi_{\psi_{s}}$ contains $\calK_{L}$,
$\pi_{\psi_{s}}(\calK_{L})=0$.
\par
Now suppose that $\lambda = \lim_{m\to\infty}\psi (P_{m}) \ne 0$. Then, $\psi (p_{j}) = \psi_{r}(p_{j})\ne 0$
for any $j$. As  $\lambda$ does not vanish, there is at least on $j$ satisfying $ \psi_{r}(p_{j})\ne 0$.
On the other hand,for a matrix unit system $p_{ij}$ of  $\calK_{L}$ satisfying $p_{ij}p_{ji}= p_i$,
the KMS condition implies
$$ \psi_{r}(p_{i}) =  \psi_{r}(p_{ij}p_{ji}) =
 e^{\epsilon_{j}-\epsilon_{i}} \psi_{r}(p_{j}), \quad \psi_{r}(p_{kl})= 0 (k\ne l)$$
 which shows that $\psi_{r}(p_{j})\ne 0$ for any $j$. 
 \par
These equation tells us that (\ref{eqn:regularKMS10000})
for $Q=AP_{m}$ ($A \in {\frak R}_L$). By taking the limit $m\to\infty$, we obtain
(\ref{eqn:regularKMS10000}) holds for any $Q\in {\frak R}_L$.
\QED

\begin{lmm} \label{lmm:AutomaticRegularKMSstate}
For any $\beta > 0$, 
if $\psi$ is a  $(\evl^{free,1}, \beta)$-KMS state of a single harmonic oscillator,
$\psi_{s} = \psi_{trivial}$ where $\psi_{trivial}$ is defined in (\ref{eqn:singularKMS10001}).
\end{lmm}
\noindent
{\bf Proof.} 
Let  $\varphi$ be a KMS state for $\evl^{free,1}$ such that
the kernel of the GNS representation for $\varphi$ contains the compact operator algebra.
Let $\{ \pi_{\varphi}(\cdot) , \Omega , \frak H \}$ be the GNS triple associated with $\varphi$.  
\par
Note that the quotient  ${\tilde{\frak R}}_1 = {\frak R}_1 \slash \calK_{1}$ is .
\par
Assuming $ \pi_{\varphi} (Q)= 0 , Q\in \calK_{L}$, 
we show $\pi_{varphi} (R(\lambda , f) = 0$. 
If $ \pi_{\varphi} (Q)= 0 , Q\in \calK_{L}$, $\varphi$ gives rise to the KMS state  
$\tilde{\varphi}$ of the quotient algebra 
${\tilde{\frak R}}  = {\frak R}_1 \slash \calK_{1}$ for the time evolution $\tilde{\alpha}_{t}^{free,1}$ induced by $\evl^{free,1}$.
\par
Let $Q$  and $R$ be entire analytic elements  in $\tilde{R}$.
Due to the KMS boundary condition and commutativity of $\tilde{R} $
, $\tilde{\varphi} (Q \sigma_{t}(R))$ is bounded  on the whole complex plane
and is entire, so $\tilde{\varphi} (Q \sigma_{t}(R))$ is a constant , 
$$\tilde{\varphi} (Q \sigma_{t}(R)) =\tilde{\varphi} (Q R) .$$
We set $Q =R = \pi_{\varphi} (f(x))$ where $f$ is a real continuous function with one variable vanishing at infinity .
($x$ is the position  operator.) 
As $\evl^{free,L}( f(x) ) = f( \cos \omega t \dot x + \sin \omega t \dot p )$ , for  $t= \pi /(2\omega)$  
$f(x) \evl^{free,L}( f(x) )$ is a compact operator. Thus,
$$\varphi (f(x)^{2}) =0 .$$
As $\Omega$ is separating for $\pi_{\varphi}(\tilde{R})^{\prime}$, 
$$\pi_{\varphi}(f(x))=0 .$$ 
It turns out
$$\pi_{\varphi} (\evl^{free,L}(f(x))) = \pi_{\varphi} (f( \cos \omega t \dot x + \sin \omega t \dot p ))$$ 
which shows that $\varphi =\psi_{trivial}$.
\QED
\bigskip
\par
The above lemma shows any KMS state for an inner perturbation of a single harmonic oscillator .
\begin{lmm} \label{sec:localKMSstate}
We consider the quantum mechanical system with one degree of freedom ${\frak R}_1$ , 
and suppose $H = p^{2} + x^{2} + V$ ($ \in \in  {\frak R}_1$ gives rise to the generator
of the time evolution $\evl$ of ${\frak R}_1$. 
If $\beta > 0$ and if $\psi$ is a  $(\evl, \beta)$-KMS state 
$$\psi =  \lambda \psi_{r} + (1-\lambda ) \psi_{trivial}$$ 
for some $\lambda$ with $0\leq \lambda \leq 1$
\end{lmm}
\noindent
{\bf Proof.} 
The perturbed  $\psi^{-V}$ is quasi-equivalent to a KMS state $\varphi$ of 
the free time evolution $\evl^{free,L}$ for which the claim of the lemma is valid. As $\psi_{trivial}^{V} =\psi_{trivial}$
we obtain our claim
\QED
\bigskip
\bigskip
\newline
We are not certain that $\psi_{s} =\psi_{trivial}$ holds for more general finite quantum systems, though,
the physical meaning of singular KMS states is clear and
in what follows, we shall consider regular KMS states. 

\begin{lmm}
For any positive integers $L < L^\prime$ and any positive function $F \in \otimes_{k \in \Lambda_{L}} L^\infty(\bfR,dx_k)$ the following estimates are valid:
\begin{eqnarray}
& & e^{-2 \beta \norm{\varphi}_\infty} {\rm Tr}_{L^\prime \backslash L}(e^{- \beta H_{L^\prime \backslash L} }) {\rm Tr}_{L}(e^{- \beta H_{L} } M_F) \leq {\rm Tr}_{L^\prime}(e^{- \beta H_{L^\prime}} M_F) \label{eq:lowerboundoftrace} \\
& & {\rm Tr}_{L^\prime}(e^{- \beta H_{L^\prime}} M_F) \leq e^{2 \beta \norm{\varphi}_\infty} {\rm Tr}_{L^\prime \backslash L}(e^{- \beta H_{L^\prime \backslash L} }) {\rm Tr}_L(e^{- \beta H_{L} } M_F)  \label{eq:upperboundoftrace}
\end{eqnarray}
where $H_{L^\prime \backslash L} = H(\Lambda_{L^\prime} \backslash \Lambda_L)$, $M_F$ is the multiplication operator of $F$ on ${\frak H}_{\Lambda_L}$ and $\norm{\varphi}_\infty$ is the supremum norm of $\varphi$.
\end{lmm}

\noindent
{\bf Proof.} 
Note that for $\beta > 0$, $e^{-\beta H_L}$ is a trace class operator on ${\frak H}_{\Lambda_L}$ and also a Hilbert-Schmidt class operator.
Thus, $e^{- \beta H_L}$ has the integral kernel $e^{- \beta H_L}(x,y)$ satisfying $\int_{\bfR^{\abs{\Lambda_L}}} \int_{\bfR^{\abs{\Lambda_L}}} \abs{e^{- \beta H_L}(x,y)}^2 dxdy < \infty$.

For $L < L^\prime$, we have
\begin{eqnarray*}
\Upsilon_{L^\prime} &=& \sum_{\Lambda \subset \Lambda_L} \Phi(\Lambda) = \sum_{k \in \Lambda_{L^\prime}} V(x_k) + \sum_{k, k+1 \in \Lambda_{L^\prime}} \varphi(x_k - x_{k+1}) \\
&\geq& \sum_{k \in \Lambda_{L^\prime}} V(x_k) + \sum_{k,k+1 \in \Lambda_{L}} \varphi(x_k -x_{k+1}) + \sum_{k,k+1 \in \Lambda_{L^\prime} \backslash \Lambda_{L}} \varphi(x_k - x_{k+1}) - 2\norm{\varphi}_\infty\\
&=& \sum_{\Lambda \subset \Lambda_{L^\prime}} \Phi(\Lambda) + \sum_{\Lambda \subset \Lambda_{L^\prime} \backslash \Lambda_{L}} \Phi(\Lambda) - 2 \norm{\varphi}_\infty \\
&=& \Upsilon_{L} + \Upsilon_{L^\prime \backslash L} - 2 \norm{\varphi}_\infty,
\end{eqnarray*}
and
\begin{equation*}
\Upsilon_{L^\prime} \leq \Upsilon_{L} + \Upsilon_{L^\prime \backslash L} + 2\norm{\varphi}_\infty.
\end{equation*}
Note that $e^{- \beta H^h_{L^\prime}}$ has the Mehler kernel $k^h_\beta(x,y) \in {\cal S}(\bfR^{2L^\prime})$,
\begin{equation*}
k_\beta^h(x,y) = \rbk{ \frac{\omega}{2 \pi \sinh(2 \omega \beta)}}^{\frac{n}{2}} \prod_{k \in \Lambda_{L^\prime}} \exp\rbk{-\frac{\omega(x_k^2 + y_k^2) {\rm coth} (2 \omega \beta) - 2 {\rm cosech} (2 \omega \beta) x_k y_k}{2}}
\end{equation*}
for $x=(x_{-L^\prime+1}, \cdots, x_{-1}, x_0, x_1, \cdots, x_{L^\prime}),y =(y_{-L^\prime+1}, \cdots, y_{L^\prime}) \in \bfR^{2L^\prime}$.
For any positive functions $f,g \in {\cal S}(\bfR^{2L^\prime})$,  we have
\begin{eqnarray*}
& & \left\langle f, (e^{- \frac{\beta H^h_{L^\prime}}{n}} e^{-\frac{\beta \Upsilon_{L^\prime}}{n}})^n g \right\rangle_{L^2} \\
&=&  \int_{{\bfR}^{2L^\prime}} f(w) \int_{\bfR^{2L^\prime}} k_{\beta/n}^h(w, z_1) e^{- \frac{\beta \Upsilon_{L^\prime}}{n}} (z_1) \int_{\bfR^{2L^\prime}} k^h_{\beta/n}(z_1,z_2) e^{-\frac{ \beta \Upsilon_{L^\prime}}{n}}(z_2) \\
& & \times \cdots \times \int_{\bfR^{2L^\prime}} k_{\beta/n}^h(z_{n-1}, z_n) e^{-\frac{\beta \Upsilon_{L^\prime}}{n}}(z_n) g(z_n) dz_n \cdots dz_2dz_1 dw \\
&\geq& e^{- 2 \beta \norm{\varphi}_\infty} \int_{{\bfR}^{2L^\prime}} f(w) \int_{\bfR^{2L^\prime}} k_{\beta/n}^h(w, z_1) e^{- \beta \frac{ \Upsilon_{L} + \Upsilon_{L^\prime \backslash L}}{n}} (z_1)  \\
& & \times \cdots \times \int_{\bfR^{2L^\prime}} k_{\beta/n}^h(z_{n-1}, z_n) e^{- \beta \frac{ \Upsilon_{L} + \Upsilon_{L^\prime \backslash L}}{n}} (z_n) g(z_n) dz_n \cdots dz_2dz_1 dw \\
&=& e^{- 2 \beta \norm{\varphi}_\infty}\left\langle f, (e^{- \frac{\beta H^h_{L^\prime}}{n}} e^{-\beta \frac{\Upsilon_{L} + \Upsilon_{L^\prime \backslash L}}{n}})^n g \right\rangle_{L^2}.
\end{eqnarray*}
Thus, we obtain
\begin{eqnarray*}
\left\langle f, e^{-\beta H_{L^\prime}} g \right\rangle_{L^2} &=& \lim_{n \rightarrow \infty} \left\langle f, (e^{- \frac{\beta H^h_{L^\prime}}{n}} e^{-\frac{\beta \Upsilon_{L^\prime}}{n}})^n g \right\rangle_{L^2} \\
&\geq& \lim_{n \rightarrow \infty} e^{-2 \beta \norm{\varphi}_\infty} \left\langle f, (e^{-\frac{\beta H^h_{L^\prime}}{n}} e^{- \frac{\beta \Upsilon_{L} + \beta \Upsilon_{L^\prime \backslash L}}{n}})^n g \right\rangle_{L^2} \\
&=& e^{- 2 \beta \norm{\varphi}_\infty} \left\langle f, e^{- \beta H_{L} - \beta H_{L^\prime \backslash L}} g \right\rangle_{L^2}.
\end{eqnarray*}
 This means that
 \begin{equation*}
 e^{-\beta H_{L^\prime}} (x,y) \geq e^{- 2 \beta \norm{\varphi}_\infty} e^{- \beta H_{L} - \beta H_{L^\prime \backslash L}} (x,y), \quad x, y \in \bfR^{2L^\prime}.
 \end{equation*}
Since $e^{- \beta H_{L^\prime}}$ is a trace class operator, the integral kernel of $e^{-\beta H_{L^\prime}}$ satisfies $\int_{R^{2L^\prime}} e^{- \beta H_{L^\prime}} (x,x) dx < \infty$. 
Thus, we obtain the following estimates for any positive function $F \in \bigotimes_{k \in \Lambda_{L}} L^\infty(\bfR)$:
\begin{eqnarray*}
{\rm Tr}_{L^\prime} (e^{-\beta H_L} M_F) &=& \int_{{\bfR}^{2L}} e^{- \beta H_L} (x,x) F(x) dx \\
&\geq& e^{-2 \beta \norm{\varphi}_\infty} \int_{{\bfR}^{2L^\prime}} e^{- \beta ( H_{L^\prime \backslash L} + H_{L} ) }(x,x) F(x) dx \\
&=& e^{-2 \beta \norm{\varphi}_\infty} {\rm Tr}_{L^\prime} (e^{- \beta H_{L^\prime \backslash L}} e^{- \beta H_{L}} M_F) \\
&=& e^{-2 \beta \norm{\varphi}_\infty } {\rm Tr}_{L^\prime \backslash L} (e^{- \beta H_{L^\prime \backslash L} }) {\rm Tr}_{L}(e^{- \beta H_{L} } M_F)
\end{eqnarray*}
and
\begin{eqnarray*}
{\rm Tr}_{L^\prime} (e^{-\beta H_{L^\prime}} M_F) &\leq& e^{2 \beta \norm{\varphi}_\infty} {\rm Tr}_{L^\prime}(e^{- \beta H_{L^\prime \backslash L}} e^{- \beta H_{L}} M_F) \\
&=& e^{2 \beta \norm{\varphi}_\infty } {\rm Tr}_{L^\prime \backslash L}(e^{- \beta H_{L^\prime \backslash L} }) {\rm Tr}_{L}(e^{- \beta H_{L} } M_F).
\end{eqnarray*}
Thus, we obtain (\ref{eq:lowerboundoftrace}) and (\ref{eq:upperboundoftrace}). \QED

\begin{pro}
\label{pro:regular01}
For any positive integers $L \leq L^\prime$ and any $F \in {\frak R}_{L}$ such that $\pi_0(F)$ is a positive multiplication operator 
on ${\frak H}_{\Lambda_{L}}$, the following estimate hold:
\begin{equation} \label{eq:estimatesofstatesofres}
e^{-4\beta\norm{\varphi}_\infty} \psi_{L} (F) \leq \psi_{L^\prime}(F) \leq e^{4 \beta \norm{\varphi}_\infty} \psi_{L} (F),
\end{equation}
where $\psi_L$ and $\psi_{L^\prime}$ are states on ${\frak R}_L$ and ${\frak R}_{L^\prime}$ defined in (\ref{eq:localKMSstate}), respectively.
\end{pro}

\noindent
{\bf Proof.} By (\ref{eq:lowerboundoftrace}) and (\ref{eq:upperboundoftrace}), we obtain the following inequalities:
\begin{eqnarray*}
& & e^{-2 \beta \norm{\varphi}_\infty } {\rm Tr}_{L \backslash L^\prime} (e^{- \beta H_{L \backslash L^\prime} }) {\rm Tr}_{L^\prime} (e^{- \beta H_{L^\prime} } M_F) \leq {\rm Tr}_L (e^{- \beta H_L} M_F),\\
& & {\rm Tr}_L (e^{- \beta H_L} M_F) \leq e^{2 \beta \norm{\varphi}_\infty } {\rm Tr}_{L \backslash L^\prime} (e^{- \beta H_{L \backslash L^\prime} }) {\rm Tr}_{L^\prime}(e^{- \beta H_{L^\prime} } M_F), \\
& & e^{-2 \beta \norm{\varphi}_\infty } {\rm Tr}_{L \backslash L^\prime} (e^{- \beta H_{L \backslash L^\prime} }) {\rm Tr}_{L^\prime}(e^{- \beta H_{L^\prime}) }) \leq {\rm Tr}_L (e^{- \beta H_L}),\\
& & {\rm Tr}_L (e^{- \beta H_L}) \leq e^{2 \beta \norm{\varphi}_\infty } {\rm Tr}_{L \backslash L^\prime}(e^{- \beta H_{L \backslash L^\prime} }) {\rm Tr}_{L^\prime} (e^{- \beta H_{L^\prime} }).
\end{eqnarray*}
Thus, we obtain (\ref{eq:estimatesofstatesofres}).\QED

Note that $\psi_L$ is also a state on ${\cal W}_L$, i.e. 
\begin{equation}
\psi_L(W) = \frac{ {\rm Tr}_L(e^{-\beta H_L} \pi_0(W)) }{ {\rm Tr}_L(e^{- \beta H_L }) }, \quad W \in {\cal W}_L. \label{eq:localKMSstateWeyl}
\end{equation}
Also, for the Weyl CCR algebra the following statement follows.
\begin{pro}
For any positive integers $L \leq L^\prime$ and any $F \in {\cal W}_{L}$ such that $\pi_0(F)$ is a positive multiplication operator 
on ${\frak H}_{\Lambda_{L}}$, the following estimate hold:
\begin{equation}
e^{-4\beta\norm{\varphi}_\infty} \psi_{L} (F) \leq \psi_{L^\prime}(F) \leq e^{4 \beta \norm{\varphi}_\infty} \psi_{L} (F) \label{eqn:555}
\end{equation}
where $\psi_L$ and $\psi_{L^\prime}$ are states on ${\cal W}_L$ and ${\cal W}_{L^\prime}$ defined in (\ref{eq:localKMSstateWeyl}), respectively.
\end{pro}

Since $e^{- \beta (p^2 + \omega^2 x^2)}$ is a trace class operator on $L^2(\bfR, dx)$ and by \cite[Proposition 5.2.27]{BratteliRobinsonII},
$e^{- \beta d\Gamma(p^2+ \omega^2 x^2)}$ is a trace class on ${\cal F}_+(L^2(\bfR, dx))$, where $d\Gamma(p^2 + \omega^2x^2)$ is the second quantization of $p^2 + \omega^2 x^2$ and ${\cal F}_+(L^2(\bfR, dx))$ is the Bose-Fock space of $L^2(\bfR,dx)$ (See also \cite[Section 5.2.1]{BratteliRobinsonII}).
Put
\begin{equation*}
\widetilde{\psi}_L := \psi_L \otimes \frac{ {\rm Tr}_{{\cal F}_+ (L^2(\bfR, dx))}(e^{- \beta H^h(\bfZ \backslash \Lambda_L)} \pi_0(\cdot) ) }{ {\rm Tr}_{{\cal F}_+ ( L^2(\bfR, dx) )}(e^{- \beta H^h(\bfZ \backslash \Lambda_L)}) },
\end{equation*}
then $\widetilde{\psi}_L$ is a regular state on ${\cal W}$ and $\widetilde{\psi}_L \restriction_{{\cal W}_L} = \psi_L$.
Thus, the regular state $\psi_L$ on ${\cal W}_L$ can extend to a regular state on ${\cal W}$.
Since ${\cal W}$ is a unital C$^*$-algebra, the family of states $\{ \widetilde{\psi}_L \}_{L \in \bfN}$ has at least one cluster point $\psi$.
Next, we show that $\psi$ is a regular state.
\bigskip

\begin{theorem}
\label{th:regularKMS}
The state $\psi$ defined in the above is a regular state on ${\cal W}$.
\end{theorem}
\noindent
{\bf Proof.} 
To show the regularity of $\psi$, we show that for $t$ in $| t |\leq \delta$
$$\widetilde{\psi}_L (Q W_{0}(t))$$ 
is equicontinous with respect to $L$
where $Q=Q(x)$ is an arbitrary essentially bounded bounded function on $\bfR^{2L}$
and $W_{0}(t)) = e^{it p_{0}}$. (We can show the continuity 
of $\lim_{L} \widetilde{\psi}_L (Q W_{0}(t))$ for the general  $W = e^{i \sum_{k=-L+1}^{L }t_{k} p_{k} }$ 
in the same way.)
\par
For simplicity of presentation, we consider the case $\omega =1$ here.
\par
Note that
\begin{eqnarray*}
&&\widetilde{\psi}_L (Q W_{0}(t)) =   
\\
&&
\frac{1}{Z_{\beta L}^{2}} 
\int\int e^{- \beta /2 H_L}(x,y) Q(x)  
\frac{e^{- \beta /2 H_L}(x+t^{(0)} ,y)}{ e^{- \beta /2 H_L}(x ,y)} e^{- \beta /2 H_L}(x ,y)dxdy 
\end{eqnarray*}
 where $Z_{\beta L}^{2}$ is the normalization constant
 $$ Z_{\beta L}^{2} = \int\int ( e^{- \beta /2 H_L}(x,y) )^{2} dxdy 
 =\int e^{- \beta H_L}(x,x)  dx .$$ 
 and $x+t^{(0)}$ is the addition of $t$ to $x$ at the component corresponding to the origin
 of the integer lattice $\bfZ$. 
 \par
For $x=(\cdots , x_{-1} ,x_{0}, x_{1}, \cdots)$ and   $y=(\cdots , y_{-1}, y_{0}, y_{1}, \cdots)$
we claim that 
\begin{equation}
\label{eqn:regular10}
e^{- c(t)}  \frac{ k_\beta^h(x_{0}+t,y_{0}) }{ k_\beta^h(x_{0},y_{0}) }  
\leq
\frac{e^{- \beta /2 H_L}(x+t^{(0)} ,y)}{ e^{- \beta /2 H_L}(x ,y)}
\leq
e^{c(t)} \frac{ k_\beta^h(x_{0}+t,y_{0}) }{ k_\beta^h(x_{0},y_{0}) }
\end{equation}
where
\begin{eqnarray*}
c (t) &=& \sup_{x_{0}} | V(x_{0}+t ) - V(x_{0})| +  
\sup_{x_{0}, x_{1}} | \varphi(x_{0} - x_{1} +t ) - \varphi(x_{0} - x_{1})|
\\
&&+\sup_{x_{0}, x_{-1}} | \varphi(x_{-1} - x_{0} +t ) - \varphi(x_{-1} - x_{0})| .
\end{eqnarray*}
$\lim_{t\to 0} c (t) = 0$ due to uniform continuity of $V$ and $\varphi$ and
this bound implies regularity.
\par
We now show (\ref{eqn:regular10}). Note the following tautological equalities holds. 
For any $n \in \bfN$,
\begin{eqnarray}
\label{eqn:regular11}
&&k_\beta^h(x_{0}+t,y_{0}) = 
\int \cdots \int  k_{\beta/ n}^h (x_{0}+t,z_{1} ) k_{\beta/ n}^h (z_{1},z_{2}) \cdots 
k_{\beta/n}^h (z_{n},y_{0}) dz_{1}\cdots dz_{n}
\nonumber
\\
&&
= \int \cdots \int  k_{\beta /n}^h (x_{0}+t,z_{1} +s_{1})k_{\beta/n}^h (z_{1}+s_{1},z_{2}+s_{2}) 
\cdots k_{\beta/n}^h (z_{n}+s_{n},y_{0}) dz_{1}\cdots dz_{n}
\nonumber
\\
\end{eqnarray}
for any constants $s_{k}$. Then, up to a multipliticave factor, $\widetilde{C}_{nt}$,
\begin{eqnarray}
\label{eqn:regular12}
&& k_{\beta /n}^h (x_{0}+t,z_{1} +s_{1})k_{\beta/n}^h (z_{1}+s_{1},z_{2}+s_{2}) 
\cdots k_{\beta/n}^h (z_{n}+s_{n},y_{0}) = 
\nonumber
\\
&&\widetilde{C}_{nt} \times
\exp [ - \frac{1}{2 \sinh (2 \beta /n)} 
\sum_{k=0}^{n} \{ \cosh(2 \beta /n)( (z_{k}+s_{k})^{2}+(z_{k+1}+s_{k+1})^{2})
\nonumber
\\ 
&&\quad\quad\quad - 2 (z_{k}+s_{k})(z_{k+1}+s_{k+1}) \} ]
\end{eqnarray}
where we set $z_{0} =x , z_{n+1} = y$.
In the exponent, we can write
\begin{eqnarray}
\label{eqn:regular13}
&&\sum_{k=0}^{n} \{
\cosh(2 \beta /n)( (z_{k}+s_{k})^{2}+(z_{k+1}+s_{k+1})^{2}) - 2 (z_{k}+s_{k})(z_{k+1}+s_{k+1})\}
\nonumber
\\
&& = \sum_{k=0}^{n} \{ \cosh(2 \beta /n)( z_{k}^{2}+ z_{k+1}^{2}) - 2 z_{k} z_{k+1} \}
\nonumber
\\
&&+ \sum_{k=0}^{n+1} A_{n, k}(s) z_{k} + \Sigma_{n} (t,x_{0},y_{0})
\end{eqnarray}
where $A_{n, k}(s)$(linear in $s_{k}$) and $ \Sigma_{n} (t,x_{0},y_{0})$(quadratic in $s_{k}$) 
are terms independent on $z_{k}$.
Now we choose the constants $s_{k}$ satisfying the condition $A_{n, k}(s) = 0$
$s_{0} =t$ $s_{n+1}=0$.
We do not need the exact form of $A_{n, k}(s)$ and  $\Sigma_{n} (s,t)$ here but
what we need are bounds $| s_{k}| \leq \widetilde{A} |t|$ independent of $n$  .
\\
Thus, we obtain
\begin{eqnarray}
\label{eqn:regular14}
&&k_\beta^h(x_{0}+t,y_{0}) = \exp\left[ - \frac{ \Sigma_{n} (t,x_{0},y_{0}) }{2 \sinh (2 \beta /n)} \right]\times
\nonumber
\\
&&
 \int \cdots \int  k_{\beta /n}^h (x_{0},z_{1})k_{\beta/n}^h (z_{1},z_{2}) 
\cdots k_{\beta/n}^h (z_{n},y_{0}) dz_{1} \cdots dz_{n} 
\end{eqnarray}
and
\begin{equation}
\label{eqn:regular15}
\lim_{n\to \infty } \exp\left[ - \frac{ \Sigma_{n} (t,x_{0},y_{0})}{2 \sinh (2 \beta /n)}\right] = 
\frac{ k_\beta^h(x_{0}+t,y_{0}) }{ k_\beta^h(x_{0},y_{0})}
\end{equation}
To show (\ref{eqn:regular10}) we apply the Trotter-Kato formula again to
\begin{equation}
\label{eqn:regular16}
e^{- \beta H^h_L}(x+t^{(0)},y) = 
\lim_{n\to\infty} (e^{- \frac{\beta}{n}H^h_L} e^{-\frac{\beta \Upsilon_L}{n}})^n (x+t^{(0)},y)
\end{equation}
We consider now
$(e^{- \frac{\beta H^h_L}{n}} e^{-\frac{\beta \Upsilon_L}{n}})^n (x+t^{(0)},y)$
at each $n$  in (\ref{eqn:regular16}). 
The integral kernel of $(e^{- \frac{\beta H^h_L}{n}} e^{-\frac{\beta \Upsilon_L}{n}})^n$ 
is an iteration of integral in which the shift $x+t^{(0)}$ of the variable affect only to
the integral associated to the particle at the origin and its nearest neighbor.   
In that  integral ,we denote  
the variable at the site $-1$ at the lattice by $z^{(-1)}_{k}$
and  that at the site $1$ at the lattice by $z^{(1)}_{k}$. Then, the contribution to the
iterated integral
from  the origin in
$(e^{- \frac{\beta H^h_L}{n}} e^{-\frac{\beta \Upsilon_L}{n}})^n (x,y)$
is
\begin{eqnarray}
\label{eqn:regular17}
\int&\cdots&\int
k_{\beta / n}^h (x_{0}, z_{1} )
\exp \left[ -\frac{\beta}{n} ( V(z_{1}) + \varphi (z_{1} - z^{(-1)}_{1})  + \varphi (z^{(1)}_{1}-z_{1})
\right]
\nonumber
\\
&\times&
k_{\beta/n}^h (z_{1},z_{2}) 
\exp \left[ -\frac{\beta}{n} ( V(z_{2}) + \varphi (z_{2} - z^{(-1)}_{2})  + \varphi (z^{(1)}_{2}-z_{2})
\right]
\nonumber
\\
&\cdots &k_{\beta/n}^h (z_{n},y_{0}) 
\exp \left[ -\frac{\beta}{n} ( V(y_{0}) + \varphi (y_{0} - y_{-1})  + \varphi (y_{1}-y_{0})
\right]  dz_{1} \cdots dz_{n}
\nonumber
\\
\end{eqnarray}
After the shift of variable $z_{k} \to z_{k}+s_{k}$ as in (\ref{eqn:regular11})
the corresponding  integral for 
$(e^{- \frac{\beta H^h_L}{n}} e^{-\frac{\beta \Upsilon_L}{n}})^n (x+t^{(0)},y)$
is
\begin{eqnarray}
\label{eqn:regular18}
\int&\cdots&\int
k_{\beta / n}^h (x_{0}, z_{1} )
\exp \left[ -\frac{\beta}{n} ( V(z_{1}) + \varphi (z_{1}-s_{1} - z^{(-1)}_{1})  
+ \varphi (z^{(1)}_{1}-z_{1}+s_{1})\right]
\nonumber
\\
&\times&
k_{\beta/n}^h (z_{1},z_{2}) 
\exp \left[ -\frac{\beta}{n} ( V(z_{2}-s_{2}) + \varphi (z_{2}-s_{2} - z^{(-1)}_{2})  
+ \varphi (z^{(1)}_{2}-z_{2}+s_{2})
\right]
\nonumber
\\
&\cdots &  dz_{1} \cdots dz_{n}  
\times \exp\left[ - \frac{ \Sigma_{n} (t,x_{0},y_{0}) }{2 \sinh (2 \beta /n)} \right]
\nonumber
\\
\end{eqnarray}
Then,
\begin{eqnarray}
&&(\ref{eqn:regular17}) \times e^{- c(\widetilde{A} t)}  
\exp\left[ - \frac{ \Sigma_{n} (t,x_{0},y_{0}) }{2 \sinh (2 \beta /n)} \right]
\nonumber
\\
&&\leq
(\ref{eqn:regular18}) \leq
(\ref{eqn:regular17}) \times
e^{c(\widetilde{A}t)}  \exp\left[ - \frac{ \Sigma_{n} (t,x_{0},y_{0}) }{2 \sinh (2 \beta /n)} \right]
\end{eqnarray}
By taking the limit $n \to \infty$ we obtain the bound (\ref{eqn:regular10}).
\par
Finally we can show the regularity of $\psi$ by using   (\ref{eqn:regular10}), 
Proposition \ref{pro:regular01}
and the Lebesgue dominated convergence theorem.
\QED


The following proposition corresponds to the Gibbs condition.
We consider the perturbation of a regular state and the automorphism $\alpha$ defined in (\ref{eq:anharmonicdynamics}).
The perturbation of an automorphism and a state on a C$^*$-algebra or a von Neumann algebra is defined in \cite[Proposition 5.4.1]{BratteliRobinsonII} and \cite[Theorem 5.4.4]{BratteliRobinsonII}.

\begin{pro} \label{sec:Gibbs}
Let $\phi$ be a regular $(\alpha, \beta)$-KMS state on ${\frak R}$, where $\alpha$ is an automorphism defined in {\rm (\ref{eq:anharmonicdynamics})} and $\beta > 0$.
Put $W(L) := \pi_0^{-1}(\Phi( \{ L, L +1 \} )) + \pi_0^{-1}(\Phi( \{ -L, -L+1 \} ))$, $L \in \bfN$.
Then $\phi$ satisfies the following condition:
\begin{equation}
\phi^{\beta W(L)} = \psi_L \otimes \widetilde{\phi} \label{eq:Gibbs condition}
\end{equation}
for all $L \in \bfN$, where $\widetilde{\phi}$ is a state over ${\frak R}_{L^c}$, $\phi^{\beta W(L)}$ is a perturbed state of $\phi$ by $\beta W(L)$. 
\end{pro}

\noindent
{\bf Proof.}
For positive integers $L < L^\prime$, let $\gamma^{L^\prime, L}_t$ be the perturbed automorphism of $\alpha^{L^\prime}_t$ by $\beta W(L)$.
Since $H_{L^\prime} - \pi_0(W(L)) = H_{L^\prime \backslash L} + H_L$ and $H_L$ and $H_{L^\prime \backslash L}$ are commute, $\gamma^{L^\prime,L}_t =\alpha_t^{L} \otimes \alpha_t^{L^\prime \backslash L}$.
The automorphism $\alpha_t^{L^\prime \backslash L}$ converges strongly to an automorphism $\alpha_t^{L^c}$ on ${\frak R}_{L^c}$ when $L^\prime \to \infty$ by Theorem \ref{sec:existenceautomorphism}.
Note that the perturbed state $\phi^{\beta W(L)}$ is a $(\gamma, \beta)$-KMS state by construction, where $\gamma_t = \alpha_t^L \otimes \alpha_t^{L^c}$.

For $0 < R \in {\frak R}_{L^c}$, we define the state $\phi_R^{\beta W(L)}$ on ${\frak R}_L$ by
\begin{equation*}
\phi_R^{\beta W(L)} (Q) = \frac{\phi^{\beta W(L)} (QR)}{\phi^{\beta W(L)} (R)}, \quad Q \in {\frak R}_L.
\end{equation*}
Note that $\phi_R^{\beta W(L)}$ is a regular state by construction and by $\gamma_t = \alpha_t^L \otimes \alpha_t^{L^c}$, $\phi_R^{\beta W(L)}$ is an $(\alpha^L, \beta)$-KMS state.
By Lemma \ref{sec:localKMSstate}, $\phi_R^{\beta W(L)} = \psi_L$.
Thus, for all $Q \in {\frak R}_L$ and $0 < R \in {\frak R}_{L^c}$
\begin{equation}
\phi^{\beta W(L)}(QR) = \psi_L(Q) \phi^{\beta W(L)}(R). \label{eq:Gibbscondition}
\end{equation}
For any self-adjoint element $R \in {\frak R}_{L^c}$ and any $\varepsilon > 0$, $R + (\norm{R} + \varepsilon) \mathbbm{1}$ is a strictly positive operator.
Then we obtain
\begin{equation*}
\phi^{\beta W(L)} (Q(R + (\norm{R} + \varepsilon)\mathbbm{1})) = \psi_L(Q) \phi^{\beta W(L)} (R + (\norm{R} + \varepsilon) \mathbbm{1}).
\end{equation*}
Since for any element $R \in {\frak R}_{L^c}$ can decompose two self-adjoint elements $R_1$ and $R_2$ such that $R = R_1 + i R_2$.
By the linearity of $\phi^{\beta W(L)}$, the equation (\ref{eq:Gibbscondition}) holds for any elements $Q \in {\frak R}_L$ and $R \in {\frak R}_{L^c}$.
Thus,
\begin{equation*}
\phi^{\beta W(L)} = \psi_L \otimes \phi^{\beta W(L)} \restriction_{{\frak R}_{L^c}}.
\end{equation*}
Put $\widetilde{\phi} = \phi^{\beta W(L)} \restriction_{{\frak R}_{L^c}}$, then we get the claim.
\QED

\begin{rmk}
For $Q \in {\frak R}$, it may not be a linear combination of  the form of $ A \bigotimes B$ for $A \in {\frak R}_L$ and 
$B \in {\frak R}_{L^c}$.
However, by Lemma \ref{sec:sdensecpt} and {\rm \cite[Theorem 4.2. (v)]{Buchholz2}}, for a regular state $\phi$ on ${\frak R}$ and any positive integer $L$, $\pi_\phi({\frak R}_L) \bigotimes \pi_\phi({\frak R}_{L^c})$ is a weakly dense sub-algebra in 
$\pi_\phi({\frak R})$.
For $Q \in {\frak R}$, there exists a positive integers $\{ L(n) \}_{n \in \bfN}$ such that $L \leq L(n)$ for any $n \in \bfN$ and a sequence $\sum_i a^{(n)}_i R^{(n)}_i \otimes K_i^{(n)}$ such that $a^{(n)}_i \in \bfC$, $R_i^{(n)} \in {\frak R}_L$, $K_i^{(n)} \in {\frak K}(\Lambda_{L(n)} \backslash \Lambda_L)$ and 
\begin{equation*}
\pi_\phi(Q) = \mathop{{\rm w}\text{-}{\rm lim}}_{n} \sum_i a^{(n)}_i \pi_\phi(A^{(n)}_i) \otimes \pi_\phi(K^{(n)}_i).
\end{equation*}
and we can defined the product state $\psi_L \otimes \phi^{\beta W(L)} \restriction_{{\frak R}_{L^c}}$
for any $Q \in {\frak R}$. 
\end{rmk}

Finally, we show uniqueness of $(\alpha, \beta)$-KMS state for $\beta > 0$ in Theorem \ref{theorem:KMSunique1}, .
Due to  Theorem \ref{th:regularKMS}, $\psi$ gives rise to a regular state on $\calW$ and hence a regular state of ${\frak R}$.

\subsection{ Proof of Theorem \ref{theorem:KMSunique1}.} 
First, we show $\psi(Q \alpha_t(R))$ is continuous in $t \in \bfR$ for any $Q, R \in {\frak R}$.
Since $\bigcup_{L \in \bfN} {\frak R}_L$ is norm dense in ${\frak R}$ and ${\frak R}_L \subset {\frak R}_{L^\prime}$ whenever $L \leq L^\prime$, we show $\psi(Q\alpha_t(R))$ is continuous in $t \in \bfR$ for any $Q, R \in {\frak R}_L$.
By Theorem \ref{sec:existenceautomorphism}, Proposition \ref{sec:continuityofregularstate} and Theorem \ref{th:regularKMS}, for any positive integer $L$ and any $Q, R \in {\frak R}_L$, $\psi(Q \alpha_t(R))$ is continuous in $t \in \bfR$.
In fact, for any $\varepsilon > 0$, there exists a positive integer $L$ such that $\norm{\alpha_t(R) - \alpha_t^L(R)} < \frac{\varepsilon}{4}$, $\abs{\psi(QR) - \psi_L(QR)} < \frac{\varepsilon}{4}$ and $\abs{\psi(Q \alpha_t^L(R)) - \psi_L(Q \alpha_t(R))} < \frac{\varepsilon}{4}$ and a $\delta > 0$ such that $\abs{\psi_L(Q \alpha^L_t(R) - Q R)} < \frac{\varepsilon}{4}$ for $\abs{t} < \delta$. Then, for $\abs{t} < \delta$
\begin{eqnarray*}
\abs{\psi(Q \alpha_t(R) - QR)} &\leq& \abs{\psi(Q \alpha_t(R) - Q \alpha_t^L(R))} + \abs{\psi(Q \alpha_t^L(R)) - \psi_L(Q \alpha_t^L(R))} \\
& & + \abs{\psi_L(Q \alpha_t^L(R) ) - \psi_L(Q R)} + \abs{\psi_L(QR) - \psi(QR)} < \varepsilon. 
\end{eqnarray*}

Next, we show that $\psi$ is an $(\alpha, \beta)$-KMS state as Definition \ref{df:KMSstate}.
Note that the following inequality are valid for any $Q, R \in {\frak R}$:
\begin{eqnarray}
\abs{\psi(Q \alpha_t(R)) - \psi_L (Q \alpha_t^L(R))} &\leq& \abs{\psi(Q \alpha_t(R)) - \psi_L(Q \alpha_t(R))} \nonumber \\
& & + \abs{\psi_L(Q \alpha_t(R)) - \psi_L(Q \alpha^L_t(R))}. \label{eq:ineqofKMS}
\end{eqnarray}
By Theorem \ref{sec:existenceautomorphism} and (\ref{eq:ineqofKMS}) and using the integral representation of an analytic function in a strip $I_\beta = \{ z \in \bfC \mid 0 < {\rm Im} z < \beta \}$ (see also the proof of \cite[Theorem 2.2]{PowersSakai}), $\psi$ is an $(\alpha, \beta)$-KMS state.

Finally, we show the uniqueness of $(\alpha, \beta)$-KMS state.
Let $\phi$ be an arbitrary extremal $(\alpha, \beta)$-KMS regular state at $\beta$.
Let $({\frak H}_\psi, \pi_\psi, \Omega_\psi)$ and $({\frak H}_\phi, \pi_\phi, \Omega_\phi)$ be the GNS representation associated with $\psi$ and $\phi$.
By $\widehat{\psi}$ and $\widehat{\phi}$, 
we denote the normal extension to the von Neumann algebra $\pi_\psi({\frak R})^{\dprime}$ and $\pi_\phi({\frak R})^{\dprime}$.

Let $\widehat{\phi}_N = \widehat{\phi}^{\beta W(N)}$, $N \in \bfN$, be the perturbed state of $\widehat{\phi}$ 
by $\beta W(N)$, where $W(N)$ is defined in Proposition \ref{sec:Gibbs}. 
Put ${\frak M} = \pi_\phi({\frak R})^{\dprime}$ and ${\frak N}_L = \pi_\phi({\frak R}_L)^{\dprime}$, $L \in \bfN$.
By Lemma \ref{sec:monotonerelentropy}, for $L \leq N$ we obtain
\begin{eqnarray*}
0 &\leq& S(\widehat{\phi_N} \restriction_{{\frak N}_L}, \widehat{\phi} \restriction_{{\frak N}_L} ) \leq S( \widehat{\phi_N}, \widehat{\phi}) = \widehat{\phi_N}(\beta W(N)) - \log \widehat{\phi_N}(\mathbbm{1}) \\
&\leq& \widehat{\phi_N}(\beta W(N)) - \widehat{\phi}(\beta W((N)) \leq 4 \abs{\beta} \norm{\varphi}_\infty.
\end{eqnarray*}
This follows from Pierls-Bogoliubov inequality:
\begin{equation*}
\log \widehat{\phi_N}(\mathbbm{1}) \geq \log e^{\widehat{\phi}(\beta W(N))} = \widehat{\phi}(\beta W(N)).
\end{equation*}
By Lemma \ref{sec:Calgrelent}, Lemma \ref{sec:restrictionstate} and Lemma \ref{sec:Gibbs}, for $L \leq N$
\begin{eqnarray*}
S(\widehat{\phi_N} \restriction_{{\frak N}_L}, \widehat{\phi} \restriction_{{\frak N}_L}) &=& S(\phi_N \restriction_{{\frak R}_L}, \phi \restriction_{{\frak R}_L} ) = S(\psi_N \restriction_{{\frak R}_L}, \phi \restriction_{{\frak R}_L}) \\ 
&=& S(\widehat{\psi_N} \restriction_{{\pi_0({\frak R}_L)^\dprime}}, \widehat{\phi} \restriction_{ \pi_0({\frak R}_L)^\dprime } ) \leq 4 \abs{\beta} \norm{\varphi}_\infty.
\end{eqnarray*}
Note that $\widehat{\psi_N} \restriction_{{\frak R}_L}$ converge to $\widehat{\psi} \restriction_{{\frak R}_L}$ in $\sigma({\cal B}({\frak H}_{\Lambda_L})_*, {\cal B}({\frak H}_{\Lambda_L}))$ topology.
By Lemma \ref{sec:wcontiofrelent}, it follows that
\begin{eqnarray*}
S(\psi \restriction_{{\frak R}_L}, \phi \restriction_{{\frak R}_L}) &=& S(\widehat{\psi} \restriction_{\pi_0({\frak R}_L)^\dprime}, \widehat{\phi} \restriction_{\pi_0({\frak R}_L)^\dprime}) \\
&\leq& \liminf_N S(\widehat{\psi_N} \restriction_{\pi_0({\frak R}_L)^\dprime}, \widehat{\phi} \restriction_{\pi_0({\frak R}_L)^\dprime}) \leq 4 \abs{\beta} \norm{\varphi}_\infty.
\end{eqnarray*}
By Lemma \ref{sec:quasicontain} and \cite[Lemma 3]{Araki1975}, we are done.
\QED
\bigskip
\bigskip
\noindent
{\bf Acknowledgment}
We would like thank Detlev Buchholz for pointing out mistakes in our first manuscript.

\end{document}